
\magnification=\magstep1 \overfullrule=0pt
\font\huge=cmr10 scaled \magstep2
\font\small=cmr8
\def\sp{\,\,\,} \def\la{\lambda} \def\eg{{\it e.g.}$\,\,$} \def\ka{\kappa}
\def\ie{{\it i.e.}$\,\,$}   \def\A{{\cal A}} \def\E{{\cal E}}
\def\R{{\cal R}} \def\J{{\cal J}} \def\p{{\cal P}} \def\O{{\cal O}}
\def\ot{\otimes}   \def\D{{\cal D}} \def\lg{\langle} \def\rg{\rangle}
\def\Ab{{\bar{\cal A}}} \def\Db{{\bar{\cal D}}} \def\Eb{{\bar{\cal E}}}
\def\GKO{[11]}  \def\CR{[5]} \def\KAC{[13]}  \def\SCH{[12]}
\def\KW{[15]} \def\GA{[6]} \def\CIZ{[4]} \def\GAN{[7]} \def\MSW{[19]}
\def\SU{[8]} \def\SUR{[10]} \def\RTW{[22]} \def\KR{[16]}
\def\KP{[14]} \def\BMW{[2]} \def\LP{[17]}  \def\RAV{[21]}
\def\WAS{[25]} \def\STA{[24]} \def\MS{[20]} \def\BER{[3]}
\def\GH{[9]} 

{\nopagenumbers \rightline{July, 1994}\vskip2cm
\centerline{{\huge On the Classification of Diagonal Coset Modular Invariants}}
\bigskip\centerline{Terry Gannon\footnote{$^*$}{\small e-mail: gannon@ihes.fr}}
\centerline{{\it Institut des Hautes Etudes Scientifiques,}}
\centerline{{\it 91440 Bures-sur-Yvette, France}}
\bigskip
\centerline{Mark A. Walton\footnote{$^\dagger$}{\small Supported in part by
NSERC. e-mail: walton@hg.uleth.ca}} \centerline{{\it Physics Department,
University of Lethbridge,}} \centerline{{\it Lethbridge, Alberta, T1K 3M4,
Canada}} \vskip2cm\centerline{{\bf Abstract}}
We relate in a novel way the modular matrices of
GKO diagonal cosets without fixed points to those of WZNW
tensor products. Using this we classify all modular invariant
partition functions of $su(3)_k\oplus su(3)_1/su(3)_{k+1}$ for all positive
integer level $k$, and $su(2)_k\oplus su(2)_\ell/su(2)_{k+\ell}$ for all $k$
and infinitely  many $\ell$ (in fact, for each $k$ a positive density of
$\ell$). Of all these
classifications, only that for $su(2)_k\oplus su(2)_1/su(2)_{k+1}$  had been
known. Our lists include many new invariants.\vfill \eject}

\pageno=1 \noindent{{\bf 1. Introduction}}\bigskip

It is believed that a large subset of all rational conformal field theories can
be generated from the Goddard-Kent-Olive (GKO) coset construction \GKO. In the
prototypical example, the minimal unitary series can be identified with the
cosets $su(2)_k\oplus su(2)_1/su(2)_{k+1}.$

This paper is concerned with the classification of modular invariant
partition functions for the diagonal GKO coset theories $g_k\oplus g_\ell/
g_{k+\ell}$, where $g_k$ is an untwisted affine algebra, at positive
integer level $k$, with horizontal subalgebra $g$. We classify what are known
as {\it physical invariants}: those modular invariants with non-negative
integer multiplicities, and a unique vacuum; no further conditions are
imposed. The connection
between this problem and the WZNW one of finding partition functions for
$g_k\oplus g_\ell\oplus g_{k+\ell}^c$ (where $g_{k+\ell}^c$ is the dual of
$g_{k+\ell}$) is well known, as is the method of constructing some of the
partition
functions for the coset by tensoring together partition functions for $g_k$,
$g_\ell$, and $g_{k+\ell}$. But by means of a simple trick the coset
classification is shown in Sect.2 to be equivalent to a small subset of the
classification for $g_k\oplus g_\ell\oplus g_{k+\ell}$ (this is more convenient
 to work with than $g_k\oplus g_\ell\oplus g_{k+\ell}^c$ -- \eg
for finding exceptionals), that can be very
easily identified (see eq. (2.8c) below). In Sect.3 we apply this to classify
the coset physical invariants for certain levels  $k,\ell$ and $g=su(2)$ --
half  of these partition functions are not listed in \eg [21,4]. Finally,
in Sects.4 and 5 we classify the $su(3)_k\oplus su(3)_1/su(3)_{k+1}$ coset
theories, and find several not included in \eg [1,5].

The classification proofs follow a general strategy developed by one of us
(T.G.) in several studies [6,7,8,10]. Modular invariants of rational
conformal field theories are of two types: (i) automorphism invariants of the
unextended chiral algebra, and (ii) those with non-identity fields coupling
to the identity \GA\ (these can be interpreted \MS\ as invariants involving
extensions of the original chiral
algebra). The first are shown (in the cases under consideration) to be
``locally outer automorphism'' invariants. That is, all fields are composed of
holomorphic and anti-holomorphic parts related by an outer automorphism
(conjugation and/or simple current) of the unextended chiral algebra. The
invariants of type (i)
are then shown to be of a simple ``global'' form, and all
possibilities are found.

The first step in classifying the type (ii) invariants is to find all fields
that can couple to the identity field, the so-called ``$\rho$-couplings''.
These possibilities are severely limited by $T$-invariance and the parity rule
of [6,22], and correspond to the possible extensions of the chiral
algebra. The well-known simple current extensions are always included, and
the automorphism invariants of these must be calculated. But other
extensions of the chiral algebra can exist, and when they do, their
automorphism invariants must all be found.

We include one table, which gives the exhaustive list of all physical
invariants for $su(3)_k\oplus su(3)_1/su(3)_{k+1}$, $\forall k$. All previously
published lists seem to miss several of these, though from a ``modern''
perspective none of these invariants should be surprising.
That our table is complete is proven in Thm.2.

\bigskip \bigskip\noindent{{\bf 2. Cosets and WZNW tensor products}}\bigskip

The point of this section is to make precise the connection between the coset
theories $g_k\oplus g_\ell/g_{k+\ell},$ without fixed points, and the WZNW
tensor product theories $g_k\oplus g_\ell \oplus g_{k+\ell}$. For $\ell=1$, we
show how one can restrict attention to $g_k\oplus g_{k+1}$. We will limit this
discussion to $g=su(n)$, but similar comments should apply to  all algebras.

\bigskip \noindent{{\it 2.1 Field identification}}\medskip

Let $P_{++}(g,k)$ be the fundamental alcove of positive highest weights of
$g_k$. We will write $k'$ for the height $k+n$. We will be
interested only in the `horizontal' parts of these weights, $\la\in P(g),$ the
weight lattice of $g$. Let $\beta_i$, $i=0,\ldots,n-1$, be the usual
fundamental weights (so $\beta_1,\ldots,\beta_{n-1}$ span $P(g))$. We will
identify a weight $\la=\sum_a\la_a\beta_a$ with its Dynkin labels $\la_a$.
So $\la\in P_{++}(g,k)$ iff each $\la_a>0$, and $\sum_{a=1}^{n-1}\la_a<k'$.
Write $\rho=(1,\ldots,1)$.

First define the following quantity, called $n$-ality \LP:
$$t_n(\la)\equiv \sum_{a=1}^{n-1}a\la_a=n\la\cdot \beta_{n-1}\ ({\rm
mod}\ n).\eqno(2.1)$$
Then $t_n(\la)\equiv 0$ iff $\la\in P(g)^*$, the root lattice of $g$.
$t_n(\la)$ is meant to generalize the {\it triality} of $su(3)$.

Let $A_k$ denote the level $k$ simple current \SCH\ =outer automorphism \BER\
which operates on  $P_{++}(g,k)$ by the formula $A_k\la=
(k'-\sum_{a=1}^{n-1}\la_a)\beta_1+\sum_{a=2}^{n-1}\la_{a-1}\beta_a$.
Usually there will be no ambiguity and we
can drop the subscript $k$ on $A_k$.

Choose any $\la,\mu\in P_{++}(g,k)$, and let $S^{(k)}$ denote the modular
$S$-matrix of $g_k$. Then:
$$\eqalignno{t_n(A^a\la)&\equiv ka+ t_n(\la)\ ({\rm mod}\ n);&(2.2a)\cr
(A^a\la)^2&\equiv \la^2-2k'a \,t_n(\la)/n+
k'{}^2a\,(n-a)/n\sp ({\rm mod}\ 2k');&(2.2b)\cr
S^{(k)}_{A^a\la,A^b\mu}&=\exp[2\pi i(b\,t_n(\la)+a\,t_n(\mu)-(a+b)\,t_n(\rho)
+kab)/n]\,S^{(k)}_{\la\mu}.&(2.2c)}$$

Define
$$\eqalignno{P_{k\ell}=\{(\la,\mu,\nu)\in & P_{++}(g,k)\times P_{++}(g,\ell)
\times P_{++}(g,k+\ell)\,|\,&\cr
&t_n(\la)+t_n(\mu)\equiv t_n(\nu)+t_n(\rho)\ ({\rm mod}\ n)\}.&(2.3)\cr}$$
For any $(\la,\mu,\nu)\in P_{k\ell}$, define $A^a(\la,\mu,\nu)=(A^a_k\la,
A^a_\ell\mu,A^a_{k+\ell}\nu)$.
We see from $(2.2a$) that $(\la,\mu,\nu)\in P_{k\ell}$ iff $A(\la,\mu,\nu)\in
P_{k\ell}$. This means
$P_{k\ell}$ is the disjoint union of $A$-orbits. Let $P^A_{k\ell}$
denote $P_{k\ell}$ after modding out by $A$, \ie{} it contains one and
only one triple $(\la,\mu,\nu)$ from each $A$-orbit $\{A^a(\la,\mu,
\nu)\}\subset P_{k\ell}$.

The characters of the coset theories are essentially the branching
functions $b_{\la\nu}^\mu$ \KAC\ (notice, however, we use positive weights
to label the $g_k$ representations, rather than non-negative ones).
We know (\eg{} [5,13]) that $b_{\la\nu}^{\mu}\ne 0$ iff $(\la,\mu,\nu)\in
P_{k\ell}$, and for $(\la,\mu,\nu)\in P_{k\ell}$,
$b_{\la\nu}^\mu=b_{\la'\nu'}^{\mu'}$ iff $(\la',\mu',\nu')=A^a(\la,
\mu,\nu)$ for some $a$.

$(\la,\mu,\nu)\in P_{k\ell}$ is called a {\it fixed point} if $A^m(\la,\mu,\nu)
=(\la,\mu,\nu)$, for some $0<m<n$. Fixed
points occur iff the greatest common divisor gcd$(n,k,\ell)>1$.
So for $n$ prime, there will be no fixed points
unless $k\equiv \ell\equiv 0$ (mod $n$).
Fixed points present certain complications, and we will restrict attention in
this paper to the simplest case, where there are  no fixed points (\eg
$\ell=1$).

Let $G$ denote the simply-connected Lie group with Lie algebra $g.$ Since the
centre of $G$ is diagonally
embedded in the centre of $G\otimes G,$ there are field identifications
[20,18,23]. That is, we must identify different triples
$(\la,\mu,\nu)\in P_{k\ell}$. For $g=su(n)$ the identifications are quite
simple (at least for unitary cosets\footnote{$^1$}{\small For nonunitary
diagonal cosets there are more field identifications, some not directly related
to the centre of $G$ \MSW.}, when no fixed points are present). The characters
of the coset model can be taken to be the branching functions,
and $(\la,\mu,\nu)$ is identified with $A(\la,\mu,\nu)$, so that the characters
are in one-to-one correspondence with the elements of $P_{k\ell}^A$.

\bigskip\noindent{{\it 2.2 The correspondence} $g_k\oplus g_\ell/
g_{k+\ell}\leftrightarrow g_k\oplus g_\ell\oplus g_{k+\ell}$}\medskip

Let $\chi^{(k)}_\la$ denote the $g_k$ character with highest weight
$\la$. As before, let its $S$ and $T$ modular matrices be denoted $S^{(k)}$
and $T^{(k)}$. For $(\la,\mu,\nu)\in P_{k\ell}^A$ define
$$\tilde{ch}_{\la\mu\nu}=\sum_{a=0}^{n-1}\,\chi_{A^a\la}^{(k)}\,
\chi_{A^a\mu}^{(\ell)}\,\chi_{A^a\nu}^{(k+\ell)*}.\eqno(2.4)$$

\noindent{{\bf Claim 1}.}\quad {\it The branching functions} $\{b_{\la\nu}^\mu
\,|\,(\la,\mu,\nu)\in P^A_{k\ell}\}$ {\it transform modularly exactly like
$\{\tilde{ch}_{\la\mu\nu}\,|\,(\la,\mu,\nu)\in P^A_{k\ell}\}$. In other
words, their $S$ and $T$ matrices are equal}.\medskip

\noindent{{\it Proof}}\quad  The $S$ and $T$ matrices for the branching
functions are computed in \KW{} and given in \CR. We find  that
$$T_{\la\mu\nu,\la'\mu'\nu'}=T^{(k)}_{\la\la'}\,T^{(\ell)}_{\mu\mu'}
\,T^{(k+\ell)*}_{\nu\nu'}.\eqno(2.5)$$
The $S$-matrix for (2.4) is no harder to calculate; from $(2.2c$) we get a
factor
$$\sum_{a=0}^{n-1} \exp[2\pi i a\{t_n(\la')+t_n(\mu')-t_n(\nu')-t_n(\rho)\}/n],
\eqno(2.6a)$$ which equals 0 unless $(\la',\mu',\nu')\in P_{k\ell}$, in
which case it equals $n$. Thus the $S$-matrices are also equal, with entries
$$S_{\la\mu\nu,\la'\mu'\nu'}=n\,S^{(k)}_{\la\la'}\,S^{(\ell)}_{\mu
\mu'}\,S^{(k+\ell)*}_{\nu\nu'}.\eqno(2.6b)$$
\qquad QED\medskip

Let $\Omega^{cos}_{k\ell}$ denote the coset commutant, \ie the space of all
{\it modular invariant combinations}
$$\sum_{(\la\mu\nu),(\la'\mu'\nu')\in P_{k\ell}^A}M_{\la\mu\nu,\la'\mu'\nu'}
\,b_{\la\nu}^\mu\,b_{\la'\nu'}^{\mu'*}.\eqno(2.7a)$$
Let ${\cal P}^{cos}_{k\ell}$ denote the set of all coset {\it physical
invariants}, \ie all functions lying in the coset commutant with all
coefficients $M_{\la\mu\nu,\la'\mu'\nu'}$ non-negative integers, and with
$M_{\rho\rho\rho,\rho\rho\rho}=1$.

Let $\Omega^{wzw}_{k\ell}$ denote the WZNW tensor product commutant, \ie
the space of all {\it modular invariant combinations}
$$\sum_{\la,\mu,\nu,\la',\mu',\nu'}
\tilde{M}_{\la\mu\nu,\la'\mu'\nu'}\,\chi_\la^{(k)}\,\chi_\mu^{(\ell)}\,
\chi_\nu^{(k+\ell)}\,\chi_{\la'}^{(k)*}\,\chi_{\mu'}^{(\ell)*}\,\chi_{\nu'}^{(
k+\ell)*},\eqno(2.7b)$$
the sum being over all $\la,\la'\in P_{++}(g,k)$ {\it etc}.
Let ${\cal P}^{wzw}_{k\ell}$ denote the set of all WZNW tensor product
{\it physical
invariants}, \ie all functions lying in $\Omega^{wzw}_{k\ell}$ with all
coefficients $\tilde{M}_{\la\mu\nu,\la'\mu'\nu'}$ non-negative integers, and
with $\tilde{M}_{\rho\rho\rho,\rho\rho\rho}=1$.

The claim allows us to define immediately a map ${\cal L}^{cw}_{k\ell}$
from $\Omega^{cos}_{k\ell}$ into $\Omega^{wzw}_{k\ell}$, as follows.
Given the coefficients $M_{\la\mu\nu,\la'\mu'\nu'}$ of some coset modular
invariant $(2.7a$), define $\tilde{M}_{\la\mu\nu,\la'\mu'\nu'}=
M_{\la\mu\nu',\la'\mu'\nu}$ (note the $\nu,\nu'$
switch\footnote{$^2$}{\small Perhaps the simplest way to see the
necessity
of this switch is to write a nonzero element of the WZNW $T$ matrix as $\exp\{
2\pi i\big((h_\la+h_\mu+h_\nu)-(h_{\la'}+h_{\mu'}+h_{\nu'})\big)\} = \exp\{
2\pi i\big((h_\la+h_\mu-h_{\nu'})-(h_{\la'}+h_{\mu'}-h_{\nu})\big)\}.$ Here
$h_{\la}+h_{\mu}+h_{\nu}$ is a WZNW conformal weight, while
$h_\la+h_\mu-h_{\nu'}$ is a coset conformal weight (mod 1).}). Then $\tilde{M}$
will define a WZNW tensor product modular invariant $(2.7b)$. Equivalently,
this amounts
to replacing each $b_{\la\nu}^\mu$ with $\tilde{ch}_{\la\mu\nu}$. From the
claim,  modular invariance is assured. This map is one-to-one; it takes the
coset
physical invariants {\it into} the WZNW tensor product physical  invariants;
however it is not {\it onto}. In particular, the image of  ${\cal
L}^{cw}_{k\ell}$ consists of all modular invariants $\tilde{Z}$ in
$\Omega^{wzw}_{k\ell}$ satisfying:

\item{(i)} $\tilde{M}_{\la\mu\nu,\la'\mu'\nu'}=0$ unless
$$t_n(\la)+t_n(\mu)-t_n(\nu')\equiv t_n(\la')+t_n(\mu')-t_n(\nu)
\equiv t_n(\rho)\quad({\rm mod}\sp n);\eqno(2.8a)$$

\item{(ii)} for all $a,b=0,\ldots,n-1$,
$$\tilde{M}_{A^a\la A^a\mu A^b\nu,A^b\la' A^b\mu'
A^a\nu'}=\tilde{M}_{\la\mu\nu,\la'\mu'\nu'}.\eqno(2.8b)$$

{}From the equation $\tilde{M}=S^{\dag}\tilde{M}S$, it
is not difficult to show that (i) holds iff (ii) holds, iff
$$\tilde{M}_{A_k\rho A_\ell \rho\rho,\rho\rho A_{k+\ell}\rho}=\tilde{M}_{\rho
\rho A_{k+\ell}\rho,A_k\rho A_\ell \rho\rho}=\tilde{M}_{\rho\rho\rho,
\rho\rho\rho}\eqno(2.8c)$$
holds, for any physical invariant $\tilde{M}$. For example, that $(2.8c)$
implies
(i) follows by looking at the $(A_k\rho A_\ell \rho\rho,\rho\rho A_{k+\ell}
\rho)$-entry
of $\tilde{M}=S^{\dag}\tilde{M}S$, and using $(2.2c)$ and the fact that
$S_{\rho\rho\rho,\la\mu\nu}>0$.

$\tilde{Z}\in \Omega^{wzw}_{k\ell}$ can be written as a sesquilinear
combination of $\tilde{ch}_{\la\mu\nu}$ over $(\la,\mu,\nu)\in P^A_{k\ell}$,
iff (2.8) is satisfied. Call this subspace $\Omega^{wzw,A}_{k\ell}$.
Then ${\cal L}^{cw}_{k\ell}$ defines an isomorphism between
$\Omega^{cos}_{k\ell}$ and $\Omega^{wzw,A}_{k\ell}$, and a bijection between
the physical invariants ${\cal P}^{cos}_{k\ell}$ and ${\cal P}^{wzw,A}_{k\ell}
={\cal P}^{wzw}_{k\ell}\cap \Omega^{wzw,A}_{k\ell}$.

The same ``switch'' of weights works in establishing a bijection between
the physical invariants (and commutants) of $g\oplus h$ and $g\oplus h^c$,
where $g,h$ are
any chiral (\eg affine) algebras and $h^c$ denotes the {\it dual} of $h$,
corresponding to modular matrices which are complex conjugates of those
of $h$. This bijection is difficult to establish by other means, since most
physical invariants $M^{gh}$ of $g\oplus h$ cannot be factorized as $M^g
\otimes M^h$. This switch is helpful in generating coset modular invariants
which would be difficult to find otherwise, though as we saw further
conditions must be satisfied in order that this
correspondence be extended to coset theories.

Incidentally, it was shown in \GA{} that the Roberts-Terao-Warner
lattice method succeeds in
generating any WZNW commutant. The mapping ${\cal L}^{cw}_{k\ell}$
tells us then that the lattice method can be successfully extended
to the diagonal cosets without fixed points, where it will also be complete.

Thus to find all physical invariants of $su(n)_k\oplus su(n)_\ell/
su(n)_{k+\ell}$ when $gcd(n,k,\ell)=1$, it suffices to find all physical
invariants of $su(n)_k
\oplus su(n)_\ell \oplus su(n)_{k+\ell}$ which satisfy $(2.8c)$.

\bigskip\noindent{{\it 2.3 Simplification when} $\ell=1$}\medskip

When one of the levels, say $\ell$, equals 1, the correspondence given
above relating the cosets to WZNW tensor products simplifies. The reason is
that
$A_1$ acts transitively on the $n$ weights in $P_{++}(g,1)$, so modding by $A$
essentially removes that factor.

When $\ell=1$, there can be no fixed points, regardless of the value of
$k$ or $n$, so the comments in this section apply to any level $k$ and any
$su(n)$.

Define $P_k=\{(\la,\mu)\in P_{++}(g,k)\times P_{++}(g,k+1)\,|\,t_n(\la)\equiv
t_n(\mu)$ (mod $n)\}$. There is an obvious bijection between $P_k$ and
$P^A_{k,1}$: identify $(\la,\mu)$ with the $A$-orbit containing $(\la,\rho,
\mu)$. Each $A$-orbit in $P_{k,1}$ contains exactly one triple of the form
$(\la,\rho,\mu)$, so we may identify the orbit, and hence the corresponding
element of $P^A_{k,1}$, with such a triple.

Define $\Omega^{wzw}_k$ to be the WZNW commutant for the direct sum
$su(n)_k\oplus su(n)_{k+1}$.

Define a map, ${\cal L}^{cw}_k$, between $\Omega^{cos}_{k,1}$
and $\Omega^{wzw}_k$ as follows: let
$$Z=\sum_{\la\mu,\kappa\nu\in P_k} M_{\la \rho\mu,\kappa \rho\nu}\,
b^\rho_{\la\mu}\,b^{\rho*}_{\kappa\nu}\eqno(2.9a)$$
be any modular invariant in $\Omega^{cos}_{k,1}$, then  define
$${\cal L}^{cw}_k(Z)=\sum_{\la\mu,\kappa\nu\in P_k} M_{\la \rho\mu,\kappa\rho
\nu}\sum_{a=0}^{n-1}
\chi^{(k)}_{A^a\la}\,\chi^{(k+1)}_{A^a\nu}\,\chi^{(k)*}_{A^a\kappa}
\,\chi^{(k+1)*}_{A^a\mu}.\eqno(2.9b)$$
(Notice again the switch: $\mu\leftrightarrow\nu.$)\medskip

\noindent{{\bf Claim 2}}.\quad ${\cal L}^{cw}_k(Z)$ {\it is a modular
invariant.}\medskip

\noindent{{\it Proof}}\quad The arguments here are similar to those used
earlier
in proving Claim 1, except that here we must transform the partition functions,
instead of simply investigating the modular behaviour of the characters
$b$ and $\tilde{ch}$ as was done there.

$T$-invariance follows as for Claim 1. $S$-invariance
is also similar to Claim 1: the calculation produces the sum
$$\sum_{a=0}^{n-1}\exp[2\pi i a\{t_n(\la)+t_n(\nu)-t_n(\kappa)-t_n(\mu)\}/n],
\eqno(2.10)$$
which equals 0 unless $t_n(\la)+t_n(\nu)\equiv t_n(\kappa)+t_n(\mu)$ (mod
$n$), \ie
unless there exists an $a$ such that both $A^a(\la,\mu),A^a(\kappa,\nu)\in
P_k$, in which case the sum equals $n$.\qquad QED\medskip

As before, ${\cal L}^{cw}_k$ is one-to-one. It also is not onto; its image
is the subspace of $\Omega^{wzw}_k$ containing those modular invariants
$\bar{Z}=\sum \bar{M}_{\la\mu,\kappa\nu}\chi_\la\chi_\mu\chi_\kappa^*
\chi_\nu^*$ satisfying:

\item{(a)} $\bar{M}_{\la\mu,\kappa\nu}=0$ unless
$$t_n(\la)+t_n(\mu)\equiv t_n(\kappa)+t_n(\nu)\quad({\rm mod}\sp n);
\eqno(2.11a)$$

\item{(b)} for all $a=0,1,\ldots,n-1$,
$$\bar{M}_{\la\mu,\kappa\nu}=\bar{M}_{A^a\la A^a\mu,A^a\kappa
A^a\nu}.\eqno(2.11b)$$

For any physical invariant $\bar M$, (a) holds iff (b) does, iff
$$\bar{M}_{\rho\rho,\rho\rho}=\bar{M}_{A\rho A\rho,A\rho A\rho}
.\eqno(2.11c)$$
As before, this follows from looking at $\bar M=S^{\dag}\bar MS$.

Let $\Omega^{wzw,A}_k$ denote this subspace; then ${\cal L}^{cw}_k$ is
an isomorphism between $\Omega^{cos}_{k,1}$ and $\Omega^{wzw,A}_k$, and
a bijection between ${\cal P}^{cos}_{k,1}$ and those physical invariants in
$\Omega^{wzw,A}_k$.

Thus to find all physical invariants of $su(n)_k\oplus su(n)_1/su(n)_{k+1}$,
it suffices to find all physical invariants of $su(n)_k\oplus su(n)_{k+1}$
which satisfy $(2.11c)$.

\bigskip\bigskip\noindent{{\bf 3. An illustration:
The cosets $su(2)_k\oplus su(2)_\ell/su(2)_{k+\ell}$}}\bigskip

As a simple illustration of the results of the previous section, we will
read off from known results for $su(2)_k\oplus su(2)_{k+1}$ and $su(2)_k
\oplus su(2)_\ell\oplus su(2)_m$ the complete list of $su(2)_k\oplus
su(2)_\ell/su(2)_{k+\ell}$ partition functions, for certain choices of
$k$ and $\ell$. In Sect.3.1 we give a new argument, based on Sect.2.3 and
\GAN, for the $\ell=1$ proof in \CIZ. In Sect.3.2 we write down all the
``obvious'' $su(2)_k\oplus su(2)_\ell\oplus su(2)_{k+\ell}$ partition functions
for general levels $k,\ell$, that satisfy (2.8c). In Sect.3.3
we prove this list is complete whenever the three greatest common divisors
$gcd(k+2,\ell+2)$, $gcd(k+2,\ell)$ and $gcd(k,\ell+2)$ are all $\le 3$.

The completeness proof for all $k$ and $\ell$ may now be within sight,
thanks to \GAN{} and recent work by Stanev \STA; here we only prove it
for the $k$, $\ell$ mentioned above.

\medskip\noindent{{\it 3.1} $su(2)_k\oplus su(2)_1/su(2)_{k+1}$}\medskip

The easiest cosets are of the form $su(2)_k\oplus su(2)_1/su(2)_{k+1}$,
and were classified in \CIZ. These constitute the minimal series.
We will give an alternative argument here.

{}From Sect.2.3 we know we must find all physical invariants
of $su(2)_k\oplus su(2)_{k+1}$ which satisfy eqs.(2.11). Using \GAN\ we can
easily
find for example all physical invariants for $su(2)_{k_1}\oplus su(2)_{k_2}$,
when $gcd(k_1+2,k_2+2)\le 3$. Here we have $k_1=k$, $k_2=k+1$,
so this gcd condition is indeed satisfied. The only physical
invariants turn out to be:

\item{(1)} ${\cal A}_k\otimes{\cal A}_{k+1}$, for all $k$;

\item{(2)} ${\cal A}_k\otimes{\cal D}_{k+1}$, for all odd $k$;

\item{(3)} ${\cal D}_k\otimes{\cal A}_{k+1}$, for all even $k$;

\item{(4)} the exceptionals ${\cal A}_{9}\otimes{\cal E}_{10}$,
${\cal E}_{10}\otimes{\cal A}_{11}$, ${\cal A}_{15}\otimes{\cal E}_{16}$,
${\cal E}_{16}\otimes{\cal A}_{17}$, ${\cal A}_{27}\otimes{\cal E}_{28}$,
and ${\cal E}_{28}\otimes{\cal A}_{29}$.

The physical invariants ${\cal A}_k$, ${\cal D}_k$ (for $k$ even), and
${\cal E}_k$ (for $k=10,16,28$) are the physical invariants
for $su(2)_k$ (their subscript
is their level); they can be found in \CIZ.

It is straightforward
to check that $(2.11c)$ is satisfied by all the invariants in (1)-(4).
Thus to each of the WZNW physical invariants given above, there
is a coset physical invariant.

This example is uncharacteristically simple: for one thing, relatively few
physical invariants of $su(n)_k\oplus su(n)_{k+1}$ or $su(n)_k\oplus
su(n)_\ell\oplus su(n)_{k+\ell}$ will be tensor products of physical
invariants of $su(n)$; for another thing, (2.8) and (2.11)
will usually be violated by most physical invariants
(\eg{} (2.8) cannot be satisfied by  any diagonal invariant).

\bigskip\noindent{{\it 3.2 $su(2)_k\oplus
su(2)_\ell\oplus su(2)_{k+\ell}$}}\medskip

Now let us write down the ``obvious'' invariants for $su(2)_k\oplus
su(2)_\ell\oplus su(2)_{k+\ell}$, for
all $k,\ell$. Some of these are included in \RAV, but some are not.
For completeness, we will include here the invariants when $k$ and $\ell$
are both even (in which case there will be fixed point), without discussing
the resolution of those fixed points.
We will begin by listing the ``simple current invariants'' \SCH. If either
$k$ or $\ell$ is odd, we find exactly 2 of these; if both $k$ and
$\ell$ are even there are exactly 6.

A simple current for $su(2)_k\oplus su(2)_\ell\oplus su(2)_{k+\ell}$
can be written as a triple $J=(J_1,J_2,J_3)$, each $J_a=0$ or 1. It
acts on a weight $\la=(\la_1,\la_2,\la_3)$ by
$$(J\la)_a=\left\{ \matrix{\la_a&{\rm if}\sp J_a=0\cr k_a+2-\la_a&{\rm
if}\sp J_a=1\cr}\right. ,\eqno(3.1a)$$
where we define $k_1=k$, $k_2=\ell$ and $k_3=k+\ell$. Define also the
quantities $J^2$ and $J\cdot \la$  by
$$J^2\equiv \sum_{a=1}^3 J_ak_a\ ({\rm mod}\ 4),\quad
J\cdot\la\equiv\sum_{a=1}^3 J_a\la_a\ ({\rm mod}\ 2).\eqno(3.1b)$$

A {\it simple current invariant} \SCH\ $M$ of $su(2)_k\oplus su(2)_\ell\oplus
su(2)_{k+\ell}$ obeys the ``local'' selection
rule
$$M_{\la\mu}\ne 0\Rightarrow \mu=J\la\quad{\rm for\ some\ } J.\eqno(3.2)$$
More precisely, any simple current $J$ with $J^2$ even can be used to define a
simple current invariant $M(J)$, in the following way:
$$[M(J)]_{\la\mu}=\sum_{m=0}^1\delta_{J^m\la,\mu}\,\delta_2[J\cdot (\la-\rho)
+mJ^2/2],\eqno(3.3a)$$
where $\delta_n[x]=1$ for $x\equiv 0$ (mod $n$), and vanishes otherwise.
The only other simple current invariant we will need we will call $M^{110}$; it
is an invariant iff $k\equiv \ell\equiv 0$ (mod 4), and is defined by
$$\bigl( M^{110} \bigr)_{\la\mu}=
\left\{ \matrix{1 &{\rm if}\sp \mu=J'\la,\quad {\rm where}\sp J'\equiv
(\la_2+1,\la_1+1,0)\ ({\rm mod}\ 2)\cr 0&{\rm otherwise}\cr}\right.
.\eqno(3.3b)$$

All simple current invariants were explicitly found for all levels and
arbitrary numbers of $su(2)$ factors, in \GAN. In our case, there will be
either 30 or 6 of them, depending on whether or not both $k$ and $\ell$ are
even. We are only interested here in those which also satisfy (2.8$c$).
The complete list of solutions is:\medskip

\item{{\bf (sc.1)}} $k,\ell$ both odd:\quad $M(111)$ and $M(111)\,M(001)$;

\item{{\bf (sc.2)}} $k$ odd, $\ell$ even: \quad $M(111)$ and $M(111)\,M(010)$;

\item{{\bf (sc.3)}} $k\equiv \ell\equiv 0$ (mod 4): \quad $M(111)$, $M(111)\,
M^{110}$, $M(111)\,M(100)$, $M(111)\,M(010)$, $M(111)\,M(001)$, and
$M(100)\,M(010)\,M(001)$;

\item{{\bf (sc.4)}} $k$ even, $\ell\equiv 2$ (mod 4):\quad $M(111)$,
$M(111)\,M(100)$, $M(111)\,M(010)$, $M(111)\,M(001)$, $M(111)\,M(100)\,
M(010)$ and $M(111)\,M(010)\,M(001)$.\medskip

In the special case where $k=\ell$, we may take the {\it conjugations} $M^c$
of each of these, defined by
$$\bigl(M^c\bigr)_{\la\mu}=M_{\la,\mu_2\mu_1\mu_3}.\eqno(3.4)$$
$M$ will obey $(2.8c$) iff $M^c$ will.

It is curious that the numbers of invariants in {{\bf (sc)}}
are precisely the numbers of (unconstrained) simple current
invariants  for $su(2)_k\oplus su(2)_\ell$.

The importance of simple current invariants (and their conjugations) is that
in all cases we know, they represent ``almost all'' of
the physical invariants. The remaining invariants are called the
{\it exceptionals}; most of these can be built up from the ${\cal E}_{10}$,
${\cal E}_{16}$ and ${\cal E}_{28}$ exceptionals of the $su(2)_k$
classification \CIZ. Those of this form which satisfy $(2.8c$) are:\medskip

\item{{\bf (e.1)}} $m=10$, $n$ odd: \quad $(\A_n\ot\E_{10}
\ot \A_{n+10})\,M(111)$ and $(\A_n\ot \A_{10-n}\ot \E_{10})\,M(111)$;

\item{{\bf (e.2)}} $m=10$, $n$ even: \quad $(\E_{10}\ot \A_{n}\ot
\A_{n+10})\,M(111)$, $(\E_{10}\ot \D_{n}\ot \A_{n+10})\,M(111)$,
$(\E_{10}\ot \A_{n}\ot \D_{n+10})\,M(111)$, $(\E_{10}\ot\D_n
\ot\D_{n+10})\,M(111)$ and the four additional invariants obtained by
interchanging the first and third components here;

\item{{\bf (e.3)}} $m=16$ or 28, $n$ odd: \quad $(\A_n\ot\E_m
\ot \A_{m+n})\,M(101)$ and $(\A_n\ot\A_{m-n}\ot\E_m)\,M(110)$;

\item{{\bf (e.4)}} $m=16$ or 28, $n$ even: \quad $(\E_m\ot \A_{n}
\ot \A_{m+n})\,M(011)$, $(\E_m\ot \D_{n}\ot \A_{m+n})\,M(011)$,
$(\A_n\ot\A_{m-n}\ot\E_m)\,M(110)$, and $(\D_n\ot\A_{m-n}\ot
\E_m)\,M(110)$;

\item{{\bf (e.5)}} $m,n\in\{10,16,28\}$ (say $m\le n$): \quad the
additional invariants in
this case are $(\E_m\ot \E_{n}\ot \A_{m+n})\,M(111)$, $(\E_{10}\ot
\E_{10}\ot \D_{20})\,M(111)$ and (if $m<n$)
$(\E_m\ot\A_{n-m}\ot\E_n)\,M(111)$.\medskip

Of course the first and second components can be interchanged, and the
conjugation (3.4) can be taken if $k=\ell$.
The lists {{\bf (sc)}} and {{\bf (e)}} will almost certainly exhaust all but
finitely many of the physical invariants for
$su(2)_k\oplus su(2)_\ell\oplus su(2)_{k+\ell}$ satisfying $(2.8c$).

The exceptionals for $su(2)_{k_1}\oplus su(2)_{k_2}$ not built from
the $\E_m$ occur at
$(k_1,k_2)=(4,4)$, (6,6), (8,8), (10,10), (2,10), (3,8), (3,28),
and (8,28) (see Ref.\ \GH). Call these $\E_{4,4}$, etc. These can be used to
construct further invariants for us:\medskip

\item{{\bf (sp.1)}} $k=\ell=4$: \quad $\E_{4,4}\ot \D_8$;

\item{{\bf (sp.2)}} $k=\ell=6$: \quad $\E_{6,6}\ot \D_{12}$;

\item{{\bf (sp.3)}} $k=\ell=8$: \quad
$\E_{8,8}\ot \D_{16}$, and $\E_{8,8}\ot \E_{16}$;

\item{{\bf (sp.4)}} $k=\ell=10$: \quad  $\E_{10,10}\ot \D_{20}$;

\item{{\bf (sp.5)}} $k=2,\ell=10$: \quad $\E_{2,10}\ot \D_{12}$;

\item{{\bf (sp.6)}} $k=2,\ell=8$:\quad $\E_{2,10}\ot \D_8$ (interchange second
and third components);

\item{{\bf (sp.7)}} $k=3,\ell=8$:\quad $(\E_{3,8}\ot \A_{11})\,M(101)$;

\item{{\bf (sp.8)}} $k=3,\ell=5$:\quad $(\E_{3,8}\ot \A_{5})\,M(101)$
(interchange second and third components);

\item{{\bf (sp.9)}} $k=3,\ell=28$:\quad $(\E_{3,28}\ot \A_{31})\,M(101)$;

\item{{\bf (sp.10)}} $k=3,\ell=25$:\quad $(\E_{3,28}\ot \A_{25})\,M(101)$
(interchange second and third components);

\item{{\bf (sp.11)}} $k=8,\ell=28$:\quad $\E_{8,28}\ot \D_{36}$;

\item{{\bf (sp.12)}} $k=8,\ell=20$:\quad $\E_{8,28}\ot \D_{20}$ (interchange
second and third components).\medskip

When $\ell=1$, we get the invariants listed in Sect.3.1 with no redundancies.
However the identity $\D_2=\A_2$ means we do get some repetition when
$\ell=2$ (the $\ell=2$ cosets include all of the $N=1$ superconformal
minimal models). We conjecture that this list gives all physical invariants of
$su(2)_k\oplus su(2)_\ell\oplus su(2)_{k+\ell}$ satisfying (2.8); in the next
subsection we prove this for certain $k,\ell$ (we have also done this for
all $k$, when $\ell=2$).

\bigskip\noindent{{\it 3.3 The small gcd case}}\medskip

The only class of $su(2)$ cosets which is classified is
that with $\ell=1$ \CIZ. In this subsection we will exploit other results
from \GAN{} which will permit us to obtain many more classifications. In
particular,
there we found (among other things) all invariants of $su(2)_{k_1}\oplus
\cdots\oplus su(2)_{k_r}$ when for each $i\neq j$, $gcd(k_i+2,k_j+2)\leq 3$.
So from this we immediately get the list of all
$su(2)_k\oplus su(2)_\ell/su(2)_{k+\ell}$ physical invariants when the
greatest common divisors $gcd(k+2,\ell+2)$, $gcd
(k+2,\ell)$, and $gcd(k,\ell+2$) are all $\leq 3$. (The relevant property
of the numbers $a\le 3$ is that the only numbers coprime to $2a$ are
$\equiv \pm 1$ (mod $2a$).) Note that for a given $k$, there are infinitely
many $\ell$ satisfying those three $gcd$ conditions -- in fact a positive
density of such $\ell$.

This argument in \GAN\ makes use of some additional properties WZNW partition
functions must satisfy \MS. We can avoid using these here, because we have only
three copies of $su(2)$, and because (2.8) holds. As usual, we will restrict
attention
here to the cosets without fixed points, \ie where at least one of $k$ and
$\ell$ is odd. With a little more work, this restriction can be lifted.

The proof of the following theorem makes use of certain useful lemmas
and techniques (\eg the {\it parity rule}) developed in several earlier papers.
 We state  these
explicitly in Sects.4 and 5, in the context of $su(3)$, in the course of
proving the more difficult Thm.2, but to avoid unnecessary duplication
we will not give their $su(2)$ translations here. For that, see in particular
\GAN.

\medskip\noindent{{\bf Theorem 1.}}\quad {\it Choose any $k,\ell$, not both
even, for which $gcd(k+2,\ell+2),\, gcd(k+2,\ell),\,
gcd(k,$ $\ell+2)$ are all $\le 3$. Then the complete list of all physical
invariants of $su(2)_k\oplus su(2)_\ell\oplus su(2)_{k+\ell}$ satisfying
(2.8) is} {\bf (sc.1)}, {\bf (sc.2)}, {\bf (e.1)} and {\bf (e.3)}.\medskip

\noindent{\it Proof} \quad The argument will follow as closely as possible the
proof of Thm.7 in \GAN. Since the detailed argument is so similar to others
(\eg analyzing the special cases $k_1=10,16,28$ reduces to arguments found
in Sect.5.3 below, and Sect.6 of \SUR), we will be somewhat sketchy here.

Exactly one of $k,\ell,k+\ell$
will be even; let $k_1$ denote this even level, and let $k_2\le k_3$ denote
the
other two, and write $k_i'=k_i+2$. Let $M$ denote any physical invariant of
$su(2)_{k_1}\oplus su(2)_{
k_2}\oplus su(2)_{k_3}$. Write $a=(a_1,a_2,a_3)$ {\it etc} for its weights,
$\rho$
for the weight $(1,1,1)$, and $S_{ab}$ for the $S$-matrix.  In particular
we first find in Sect.6 of \GAN\ the consequences of the so-called
parity rule (in fact the reason for our $gcd$ conditions is precisely to
maximize the effectiveness of this parity rule).
We get the following selection rules:
$$\eqalignno{M_{ab}\ne 0 &\sp \Rightarrow \sp b_2\in\{a_2,k_2'-a_2\},\sp
b_3\in\{a_3,k_3'-a_3\};&(3.5a)\cr
M_{ab}\ne 0&\sp \Rightarrow \sp a_1^2\equiv b_1^2\sp({\rm mod}\ k_1');&(3.5b)
\cr
M_{ab}\ne 0,\ a_1=1,\ k_1\ne 10,28&\sp\Rightarrow \sp
b_1\in\{1,k_1'-1\};&(3.5c)\cr
M_{ab}\ne 0,\ a_1=1,\ k_1=10&\sp\Rightarrow \sp b_1\in\{1,5,7,11\};&(3.5d)\cr
M_{ab}\ne 0,\ a_1=1,\ k_1=28&\sp\Rightarrow \sp b_1\in\{1,11,19,29\}.&(3.5e)
\cr}$$
Eq.$(3.5a$) is Lemma 3(a) in \GAN, $(3.5b$) is T-invariance using $(3.5a$),
and $(3.5c$)-$(3.5e)$ are Lemma 3(b) in \GAN, using $(3.5b)$.

Let ${\cal R}_L=\{a\,|\,M_{a\rho}\ne 0\}$, ${\cal R}_R=\{b\,|\,M_{\rho b}\ne
0\}$. Write ${\cal J}_L$ for the set of all simple currents $J$ such that
$J\rho\in {\cal R}_L$; define ${\cal J}_R$ similarly. Much is known about
the sets ${\cal R}_{L,R}$ and ${\cal J}_{L,R}$ -- see \eg \GAN, \SUR, and
Lemmas 4 and 5 in Sect.5 of this paper. For instance the cardinalities
$\|{\cal J}_L\|$ and $\|{\cal J}_R\|$ must be equal, and ${\cal J}_{L,R}$
both are groups (under component-wise addition mod 2). Also each norm $J^2$
(see $(3.1b$)) in ${\cal J}_{L,R}$ must be 0, and $M_{Ja,J'b}=M_{ab}$
$\forall J\in{\cal J}_L,J'\in{\cal J}_R$. Also, $S_{a,Jb}=-1^{J\cdot(a-\rho)}
S_{ab}$, and for $J\in{\cal J}_L$, $J\cdot(a-\rho)\equiv 0$ if $M_{ab}\ne
0$ for some $b$. Recall the definition
of simple current invariant, given in (3.2); all simple current
invariants for $su(2)_{\ell_1}\oplus\cdots\oplus su(2)_{\ell_r}$ are
explicitly known, and in our case will lie in {{\bf (sc.1)}} or {{\bf
(sc.2)}}.

\noindent{{\it Case 1}}: Consider first the case where all $a\in {\cal R}_L\cup
{\cal
R}_R$ satisfy $a_1\in\{1,k_1'-1\}$ -- \ie ${\cal J}_L\rho={\cal R}_L$ and
${\cal J}_R\rho={\cal R}_R$. By $(3.5c)$, this is the case most of the time.
$a$ is called a {\it fixed point} of ${\cal J}_L$ if $Ja=a$ for some $J\in
{\cal J}_L$, $J\ne 0$. Note that ${\cal J}_L$ has a fixed point iff $(1,0,0)\in
{\cal J}_L$, in which case $a$ is a fixed point iff $a_1=k_1'/2$. Similar
comments apply to ${\cal J}_R$.

Suppose $M_{ab}\ne 0$, and $a$ is not a fixed point of ${\cal J}_L$, and
$b$ is not one of ${\cal J}_R$. Without loss of generality suppose $S_{a\rho}
\le S_{b\rho}$. Then from the various results about ${\cal J}_{L,R}$ and
$S$ given earlier, we get
$$\|{\cal J}_R\|\,M_{ab}\,S_{b\rho}\le (MS)_{a\rho}=(SM)_{a\rho}=\|{\cal
J}_L\|\,S_{a\rho}.\eqno(3.6)$$
Since $\|{\cal J}_L\|=\|{\cal J}_R\|$, this tells us that $M_{ad}=1$ iff $d\in
{\cal J}_Rb$, otherwise it equals 0. It also tells us that $S_{b\rho}=
S_{a\rho}$, \ie (using $(3.5a)$) $b_1\in\{a_1,k_1'-a_1\}$. Similarly, we get
that $M_{cb}=1$ iff $c\in {\cal J}_La$, otherwise it equals 0.
Thus if there are no fixed points,
$M$ must satisfy (3.2) and we are done. We may assume then that
$k_1\equiv 0$ (mod 4)
and $(1,0,0)\in{\cal J}_R$, say. We may also assume $k_1>4$, since $(3.5a),
(3.5b)$ force (3.2) if $k_1=4$.

Now put $x=(3,1,1)$ and choose $y$ so that $M_{xy}\ne 0$. $x$ is not a
fixed point of ${\cal J}_L$; for now assume
$y$ is not one of ${\cal J}_R$. Then by the previous paragraph,
$y_1\in\{3,k_1'-3\}$. Using this, we can prove (3.2) holds for all $a,b$,
as follows. Suppose $M_{ab}\ne 0$, where $a_1\ne k_1'/2$, but
$b$ is a fixed point of ${\cal J}_R$. Then $SM=MS$ applied to $(a,\rho)$ and
$(a,y)$ gives us
$$1=2\sin(\pi a_1/k_1'),\quad \pm 1=2\sin(3\pi a_1/k_1'), \eqno(3.7)$$
which are incompatible. This implies $M$ is a simple current
invariant, and we are done.

If instead $y$ defined in the previous paragraph is a fixed
point of ${\cal J}_R$, then $y_1=k_1'/2$, and $(3.5b$) forces $k_1=16$. Suppose
$M_{ab}\ne 0$. Then $(3.5b)$ tells us that if $a_1\not\in\{3,9,15\}$, then
$b_1\in\{a_1,k_1'-a_1\}$, and if $a_1\in\{3,9,15\}$ then $b_1\in\{3,9,15\}$
(incidently, by Lemma 4(b) this forces $(1,0,0)\in{\cal J}_L$). Suppose
$a_1=3$ but $b_1\ne 9$; then from $(SM)_{xb}=(MS)_{xb}$ we get $2S_{ax}=\pm
S_{ay}$, \ie $2\sin(9\pi/18)=\pm\sin(27\pi/18)$, which is impossible.

This proves (using Lemmas 4(b) and 5(a)) that $a_1$ and $b_1$ are related to
each other independently of the values of $a_2,a_3,b_2,b_3$. The reverse can
also be seen to hold, using T-invariance, $(3.5a$), and (2.8). Thus $M$ will
be the tensor product of $\E_{16}$ with some simple current invariant of
$su(2)_{k_2}\oplus su(2)_{k_3}$, and will be listed in {{\bf (e.3)}}.

\noindent{{\it Case 2}:} By $(3.5c$), Case 1 handled all values of $k_1$ except
10 and 28. These exceptional levels use familiar arguments (see \eg
Sect.5.3) and we will not repeat them here. The idea is to first find
the possibilities for the values of $M_{\rho b}$ and $M_{a\rho}$ (this
is done in the proof of Thm.7 in \GAN), and then use Lemmas 4(b) and 5
and modular invariance (particularly the relation $SM=MS$) to find the
other values of $M$. \qquad QED

\bigskip\bigskip\noindent{{\bf 4. The physical invariants of }
$su(3)_k\oplus su(3)_1/su(3)_{k+1}$} \bigskip

The last section was an illustration of carrying over known WZNW
classifications directly to the GKO classification. Unfortunately, there
are few examples of WZNW classifications, even for simple $g$. One of
these is $g=su(3)$. Our goal in this section and the next is to prove
the following theorem, which we learned in Sect.2.3 solves the
classification  problem for the cosets $su(3)_k\oplus su(3)_1/su(3)_{k+1}$,
$\forall k$:

\medskip \noindent{\bf Theorem 2}.\quad {\it The list of all physical
invariants of $su(3)_k\oplus su(3)_{k+1}$ satisfying (2.11c) is given
in the Table.}\medskip

In the Table and throughout the remainder of this paper, $\Ab_k$, $\Db_k$ and
$\Eb_k$ denote the physical invariants
of $su(3)_k$, and are explicitly given in \SU. The simple current
invariants $\bar{M}(A_kA_{k+1})$ are defined in (4.4), and their
conjugation ${}^c[\bar{M}(A_kA_{k+1})]^c$ is defined in $(4.10a)$. The
invariants
of $su(3)_k\oplus su(3)_1/su(3)_{k+1}$ were discussed in [1,5], but
their lists are very incomplete.

Sections 4 and 5 are devoted to proving Thm.2. Our task in this
section is to  find all physical invariants which are automorphisms of the
unextended chiral algebra, but we will begin with some general observations
which will also be useful in the following section.

\bigskip\noindent{{\it 4.1 General comments}}\medskip

As usual write $\la=(\la_1,\la_2)$ for an $su(3)$ weight, $k'=k+3$ for the
height, $t(\la)\equiv\la_1-\la_2$ (mod 3), $A_k$ for the simple current
taking $\la$ to $A_k\la=(k'-\la_1-\la_2,\la_1)$ and $C$ for the conjugation
$C\la=(\la_2,\la_1)$. Write $\p(k)$ for
$P_{++}(su(3),k)\times P_{++}(su(3),k+1)$.
Together, $A_k$ (which is order 3) and $C$ (which is order 2) generate the
order 6 group of outer automorphisms ${\cal O}^k$ of affine $su(3)$. Let
${\cal O}^k\la$
denote the orbit $\{C^aA^b_k\la\,|\,a=0,1;\sp b=0,1,2\}$ of $\la$ by this
group. Also, let ${\cal O}^k_0\la$ be the orbit
$\{A^b_k\la\,|\,b=0,1,2\}$ of $\la$ by the simple currents. For example,
${\cal O}^k\rho={\cal O}^k_0\rho=\{(1,1),\,(k+1,1),\,(1,k+1)\}$.

What we need to consider is $su(3)_k\oplus su(3)_{k+1}$. Note first
that the heights $k'$ and $k'+1$ are coprime, so the two
summands are {\it almost} independent. For example, $T$-invariance implies
that if $M_{\la\mu,\ka\nu}\neq 0$ for any invariant $M$, then
$$3\la^2\equiv 3\ka^2\sp({\rm mod}\sp 2k');\quad 3\mu^2\equiv 3\nu^2\ ({\rm
mod}\ 2k'+2).\eqno(4.1)$$

Also very important is
the parity rule of \GA(=the arithmetical symmetry of \RTW). For $su(3)_k
\oplus su(3)_{k+1}$ it reduces to the following. Let $\la\in P(su(3))$.
There exists a unique weight, call it $[\la]_k$, lying in both $P_{+}(su(3)
,k+3)$ and the orbit of $\la$ by the affine Weyl group. If $[\la]_k$ also lies
 in
$P_{++}(su(3),k)$, then there exists a {\it unique} (finite) Weyl automorphism
$\omega$ and vector $\alpha\in P(su(3))^*,$ the root lattice of $su(3),$ such
that
$[\la]_k=\omega(\la+ k'\alpha)$; in this case we can define the {\it parity}
$p_k(\la)$ of $\la$ to be the parity $\epsilon(\omega)=$det$(\omega$). The
parity rule tells us that for any invariant $M$, any integer $\ell$ coprime to
$3k'(k'+1)$, and any $(\la\mu),(\ka\nu)\in \p(k)$, we have
$$M_{\la\mu,\ka\nu}=p_k(\ell\la)\,p_{k+1}(\ell\mu)\,p_k(\ell\ka)\,p_{k+1}(
\ell\nu)\,M_{\la'\mu',\ka'\nu'},\eqno(4.2a)$$
where $\la'=[\ell\la]_k$, $\mu'=[\ell\mu]_{k+1}$, $\ka'= [\ell\ka]_k$ and
$\nu'=[\ell\nu]_{k+1}$.

Because $k'$ and $k'+1$ are coprime, this simplifies a little. Given any
$\ell_1$ and $\ell_2$, there will be an $\ell$ satisfying
$$\ell\equiv \ell_1\sp({\rm mod}\sp 3k')\sp {\rm and}\sp \ell\equiv\ell_2\sp
({\rm mod}\sp 3(k'+1)),\eqno(4.2b)$$
iff $\ell_1\equiv \ell_2$ (mod 3). Let $\ell_1$ be any integer coprime to
$3k'$, and choose $\ell_2=\pm 1$ so that $\ell_1\equiv\ell_2$ (mod 3). Find
any $\ell$ satisfying $(4.2b)$.
It is easy to see that, for any $\mu,\nu\in P_{++}(su(3),k+1)$, $p_{k+1}(\ell
\mu)=\ell_2=p_{k+1}(\ell\nu)$.
Then $(4.2a$) implies the following equation, familiar from \SU.
Let $M$ be any
physical invariant, and suppose $M_{\la\mu,\ka\nu}\ne 0$. By $\{x\}$ we
mean the unique number $y$ satisfying both $0\leq y<k'$ and $y\equiv x$ (mod
$k'$). Then for all integers $\ell_1$ coprime to $3k'$,
$$\{\ell_1\la_1\}+\{\ell_1\la_2\}<k'\sp{\rm iff}\sp
\{\ell_1\ka_1\}+\{\ell_1\ka_2\}<k';\eqno(4.3)$$
a similar statement holds for $\mu$ and $\nu$, with $k'$ replaced with $k'+1$.
We will later need other consequences of the parity rule $(4.2a$).

Two consequences of $(4.1)$ and $(4.3)$ have already been drawn in the
literature. In \KR{}
in a completely different context (this paper was ``discovered'' and
brought into the context of modular invariants by \RTW), it was proved
that:

\medskip
\noindent{{\bf Lemma 1}}.\quad{\it When $k'$ is coprime to 6, $\la$ and $\ka$
satisfy (4.3) iff} $\ka\in {\cal O}^k\la$. \medskip

Of course an analogous statement holds for $\mu,\nu$ and $k'+1$. Eq.(4.1) was
not used
in this derivation. This remarkable result has the flaw that it only holds
for some $k$. When $k'$ and 6 are not coprime, the
situation becomes considerably more
complicated, and no classification of the solutions $\la,\ka$ to (4.3)
is known for general $k$ (it would likely be very messy and probably
useless). However, if we also use (4.1) and restrict ourselves to the
special case $\ka=\rho$, all $\la$ solving (4.3) {\it are} known, for {\it
all} $k$ (this is Prop.1 in \SUR):

\medskip
\noindent{{\bf Lemma 2}}. \quad {\it The set of all solutions $\la$ to (4.1)
and (4.3), for $\ka=\rho$, is:

\item{(a)} for $k\equiv 0,2,3$ (mod 4), $k\neq 15$:
$\la\in{\cal O}^k_0\rho$;

\item{(b)} {\it for} $k\equiv 1$ (mod 4), $k\neq 9,21,57$:
$\la\in{\cal O}^k_0\rho\,\cup\,{\cal O}^k_0(\,(k+1)/2,(k+1)/2\,)$;

\item{(c)} {\it for $k=9,15,21,57$, respectively, $\la$ lies in}
$$\eqalign{&{\cal O}^9_0\rho\,\cup\,{\cal O}^9(3,3)_0\,\cup\,{\cal O}^9_0
(5,5),\cr&{\cal O}^{15}_0\rho\,\cup\,{\cal O}^{15}(1,4),\cr
&{\cal O}^{21}_0\rho\cup\,{\cal O}^{21}_0(5,5)\,
\cup\,{\cal O}^{21}_0(7,7)\,\cup\,{\cal O}^{21}_0(11,11),\cr
&{\cal O}^{57}_0\rho\,\cup\,{\cal O}^{57}_0(11,11)\,
\cup\,{\cal O}^{57}_0(19,19)\,\cup\,{\cal O}^{57}_0(29,29).\cr}$$}

Some useful results concerning the modular $S$-matrix $S^{(k)}$ of $su(3)_k$
are Claims 1, 2, 8 in \SUR:

\medskip\noindent{{\bf Lemma 3}}. \quad (a) {\it Suppose $S^{(k)}_{\la\rho}=
S^{(k)}_{\ka\rho}$ and $\{\la_1,\la_2,k'-\la_1-\la_2\}\cap\{\ka_1,\ka_2,k'-
\ka_1-\ka_2\}\ne\{\}$. Then} $\ka\in\O^k\la$.

\item{(b)} {\it For any $\la$, $a,b$,
$S^{(k)}_{(1,2), C^aA^b\la}=S^{(k)}_{(1,2),\la}$ iff} $C^aA^b\la=\la$.

\item{(c)} {\it For any $\la$, $a,b$,
$S^{(k)}_{(1,4), C^aA^b\la}=S^{(k)}_{(1,4),\la}$ iff} $C^a\la\in\{\la,
A\la,A^2\la\}$.

\medskip Finally, we will define \SCH\ the physical invariant $\bar{M}(J)$ of
$su(
3)_k\oplus su(3)_{k+1}$ associated with the simple current $J=A_k^aA_{k+1}^b$,
for $a,b$ not both divisible by 3 (compare (3.$3a$)):
$$[\bar{M}(J)]_{\la\mu,\ka\nu}=\sum_{m=0}^2\delta_{J^m(\la\mu),\ka\nu}\,
\delta_3[a\,t(\la)+b\,t(\mu)-m(ka^2+(k+1)b^2)].\eqno(4.4)$$

\medskip\noindent{{\it 4.2 The automorphism invariants of} $su(3)_k\oplus
su(3)_{k+1}$}\medskip

A natural first step is to
find all automorphism(=permutation) invariants corresponding to that direct
sum.
The hardest part of this (Claim 3 below) will be to show that ``locally'' the
automorphism invariant acts like an outer automorphism.

By an {\it automorphism invariant} we mean a physical invariant $M=M^\sigma$
with
$$(M^\sigma)_{\la\mu,\ka\nu}=\delta_{\ka\nu,\sigma(\la\mu)}$$
where $\sigma$ is a permutation of $\p(k)$.
The entries of $M^\sigma$ will be a bunch of 0's and 1's, with exactly
one `1' on each row and column. Of course, $M^\sigma_{\rho\rho,\rho\rho}=1$
means that $\sigma(\rho\rho)=(\rho\rho)$.

Commutation of $M^\sigma$ with $S^{(k,k+1)}=S^{(k)}\ot S^{(k+1)}$ means that
$$\eqalignno{S^{(k)}_{\la\ka}\,S^{(k+1)}_{\mu\nu}&=
S^{(k)}_{\la'\ka'}\,S^{(k+1)}_{\mu'\nu'},&(4.5a)\cr
N^{(k)}_{\la\ka\alpha}\,N^{(k+1)}_{\mu\nu\beta}&=
N^{(k)}_{\la'\ka'\alpha'}\,N^{(k+1)}_{\mu'\nu'\beta'},&(4.5b)\cr}$$
where $(\la\mu),(\ka\nu),(\alpha\beta)\in\p(k)$, where $\sigma(\la\mu)=(\la'
\mu')$, $\sigma(\ka\nu)=(\ka'\nu')$ and $\sigma(\alpha\beta)=(\alpha'\beta')$,
and where $N^{(k)}$ are the fusion
coefficients for $su(3)_k$ (similarly for $su(3)_{k+1}$).
By Verlinde's formula, we know $(4.5a)$ implies $(4.5b)$.

The entries of
$S^{(k)}$ are given by the Kac-Peterson formula \KP, while those of
$N^{(k)}$ (for $su(3)$) were computed in \BMW.
Two important special cases are:
$$\eqalignno{S^{(k)}_{\rho\la}=&{2\over \sqrt{3}k'}\{s(2\la_1)+s(2\la_2)+
s(2\la_0)\}={8\over \sqrt{3}k'}\,s(\la_1)\,s(\la_2)\,s(\la_0)>0,&(4.6a)\cr
N^{(k)}_{\la\la\la}=&\,{\rm min}_{i=0,1,2}\{\la_i\},&(4.6b)\cr}$$
where $\la_0=k'-\la_1-\la_2$ and $s(x)=\sin(\pi x/k')$. From (4.6$a$) we see
that $\sqrt{3}\,S^{(k)}_{\rho\la}$ is a polynomial over {\bf Q} in $\xi_{k'}=
\exp(2\pi i/k')$. But we know (see \eg \WAS) that the roots of unity
$\xi_m$ and $\xi_n$ are {\it algebraically independent} over {\bf Q},
whenever $m$ and $n$ are coprime. Using
$(4.5a$) and the fact that $k'$ and $k'+1$ are coprime, what this means is that
$$\sigma(\la,\mu)=(\la',\mu')\sp\Rightarrow\sp {S^{(k)}_{\rho\la}\over
S^{(k)}_{\rho\la'}}\sp =\sp {S^{(k+1)}_{\rho\mu'}\over S^{(k+1)}_{\rho\mu}}\in
{\bf Q}.\eqno(4.6c)$$
Another convenient formula for the S-matrix of $su(3)_k$ is, for any weights
$\la,\ka$,
$$ S^{(k)}_{\la,C\ka}=(S^{(k)}_{\la\ka})^*.\eqno(4.6d)$$

Our first task will be to prove:

\medskip
\noindent{{\bf Claim 3}}.\quad {\it Any automorphism invariant $M^\sigma$ for
$su(3)_k\oplus su(3)_{k+1}$ acts {\it locally} like an outer automorphism.
In other words, choose any $(\la\mu)\in \p(k)$
and write $\sigma(\la\mu)=(\la'\mu')$; then both $\la'\in {\cal O}^k\la$ and
$\mu'\in {\cal O}^{k+1}\mu$.}\medskip

\noindent{{\it Proof}} \quad
Suppose $(\la\mu)$ has $\la'\in{\cal O}^k\la$. Then
eqs.(4.5) and (4.6) tell us that
$$S_{\rho\mu}^{(k+1)}=S_{\rho\mu'}^{(k+1)},\quad N_{\mu\mu\mu}^{(k+1)}
=N_{\mu'\mu'\mu'}^{(k+1)}.\eqno(4.7)$$
Lemma 3(a) then tells us that
$\mu'\in\O^{k+1}\mu$. Thus $\la'\in\O^k\la$ iff $\mu'\in\O^{k+1}\mu$.

But for $k\equiv 2,4$ (mod 6) Lemma 1 tells us
that $\la'\in {\cal O}^k\la$, and for $k\equiv 1,3$ (mod 6) it tells us that
$\mu'\in{\cal O}^{k+1}\mu$. This concludes the proof of our Claim for
$k\equiv 1,2,3,4$ (mod 6). It remains to consider $k\equiv 0,5,6,11$ (mod 12).
For this purpose, we will make use of the following important fact:

For any $(\la\mu),(\ka\nu)\in\p(k)$ and integers $a,b$,
$$\sigma(\rho\mu)=(A^a\rho,\nu)\sp{\rm and}\sp\sigma(\la\rho)
=(\ka,A^b\rho)\Rightarrow\sigma(\la\mu)=(A^a\ka,A^b\nu).\eqno(4.8a)$$
Eq.(4.8$a$) holds for any level $k$, and analogues will hold for all algebras
$g$. To see it, write
$\sigma(\la\mu)=(\la'\mu')$; then $\la',\mu'$ are the unique weights
satisfying
$$S^{(k)}_{\alpha\la}\,S^{(k+1)}_{\beta\mu}=S^{(k)}_{\alpha'\la'}\,
    S^{(k+1)}_{\beta'\mu'}\eqno(4.8b)$$
for all $(\alpha\beta)\in\p(k)$, where $\sigma(\alpha\beta)=(\alpha'\beta')$.
 (This follows \eg from orthogonality of the rows of $S^{(k,k+1)}$.)
Using $(2.2c)$, the two hypotheses in (4.8$a$) tell us
$$S^{(k)}_{\alpha\rho}\,S^{(k+1)}_{\beta\mu}=\omega^{at(\alpha')}\,
    S^{(k)}_{\alpha'\rho}\,S^{(k+1)}_{\beta'\mu'},\quad
    S^{(k)}_{\alpha\la}\,S^{(k+1)}_{\beta\rho}=\omega^{bt(\beta')}
    S^{(k)}_{\alpha'\la'}\,S^{(k+1)}_{\beta'\rho},\eqno(4.8c)$$
where $\omega=\exp(2\pi i/3)$. Multiplying these and dividing by the
corresponding
equation for $\sigma(\rho\rho)=(\rho\rho)$ gives us (4.8$b$) with $(\la'\mu')
=(A^a\ka,A^b\nu)$, completing the proof of $(4.8a)$.

One immediate consequence of $(4.8a)$ is that if we know that Claim 3 holds
for  $(\la\rho)$,  and $(\rho\mu)$, then it holds for $(\la\mu)$.

Look first at $k\equiv 11$ (mod 12). Choose any $(\la\mu)\in\p(k)$, and write
$\sigma(\la\rho)=(\la'\rho')$ and $\sigma(\rho\mu)=(\rho''\mu')$. By Lemma
2(a) we read that $\rho'\in{\cal O}^k_0\rho$ and $\rho''\in{\cal O}^{k+1}_0
\rho$.
By the conclusion of the paragraph containing (4.7), this forces both $\la'\in
{\cal O}^k\la$ and $\mu'\in{\cal O}^{k+1}\mu$. By the previous paragraph,
this concludes the proof of Claim 3 for these $k$.

The identical argument and conclusion applies to $k\equiv 6$ (mod 12).
For $k\equiv 0,5$ (mod 12), respectively, we have to consider the
additional possibilities that $\rho'\in{\cal O}^{k+1}_0(\,(k+2)/2,(k+2)/2\,)$
and $\rho''\in{\cal O}^k_0(\,(k+1)/2,(k+1)/2\,)$. Consider without loss of
generality the latter case. Then (4.6$c$) becomes the statement that
$${S^{(k)}_{\rho\rho}\over{S^{(k)}_{\rho\rho''}}}=
{\sin(\pi/k')\,\sin(\pi/k')\,\sin(\pi(k+1)/k')\over
\sin(\pi(k+1)/2k')\,\sin(\pi(k+1)/2k')\,\sin(2\pi/k')}={2\over
1+\cos(2\pi/k')}-1 \eqno(4.9a)$$
is a rational number. In other words, $k'$ must satisfy
$$\cos(2\pi/k')\in{\bf Q}.\eqno(4.9b)$$

For $n>2$, the degree $[{\bf Q}(\cos(2\pi/n)):{\bf Q}]$ of the extension of
${\bf Q}$ by $\cos(2\pi/n)$ equals $[{\bf Q}(\exp(2\pi i/n)):{\bf Q}]/
[{\bf Q}(\exp(2\pi i/n)):{\bf Q}(\cos(2\pi/n))]=\phi(n)/2$, where $\phi(n)$
is the number of $1\le\ell<n$ coprime to $n$. Therefore
$(4.9b)$ holds iff $\phi(k')\le 2$, \ie iff $k'=1,2,3,4,6$.\footnote{${}^3
$}{{\small
We thank Antoine Coste for this proof that (4.9b) implies k'=1,2,3,4,6.}}

But none of these
solutions to $(4.9b$) satisfy $k'=k+3\equiv 8$ (mod 12). This contradiction
means that $\rho''$ must lie in
${\cal O}^k\rho$. The familiar argument
now forces $\mu'\in{\cal O}^{k+1}\mu$, completing the proof of Claim 3
for $k\equiv 5$ (mod 12). The argument for $k\equiv 0$ (mod 12) is identical.
\qquad QED\medskip

Given any physical invariant $M$, define the conjugations $M^c,{}^cM,{}^cM^c$
by
$$(M^c)_{\la\mu,\ka\nu}=M_{\la\mu,\ka C\nu},\sp
({}^cM)_{\la\mu,\ka\nu}=M_{\la\mu,C\ka\nu},\sp
({}^cM^c)_{\la\mu,\ka\nu}=M_{\la\mu,C\ka C\nu}.\eqno(4.10a)$$
Then each of these conjugations will also be a physical invariant.

\medskip\noindent{{\bf Claim 4}}. \quad $M$ {\it is an automorphism invariant
of $su(3)_k\oplus su(3)_{k+1}$ iff one of the 4 conjugations $(4.10a)$ of $M$,
call it $M^{\sigma'}$, satisfies}:
$$\forall(\la\mu)\in\p(k),\sp\sigma'(\la\mu)=(A^{a\,t(\la)+b\,t(\mu)}\la,
A^{c\,t(\la)+d\,t(\mu)}\mu),\eqno(4.10b)$$
{\it for fixed $(a,b,c,d)$, independent of $\la,\mu$. Moreover,}
\item{(a)} {\it for} $k\equiv 0$ (mod 3), $(a,b,c,d)\in\{(0,0,0,0),\,(1,\pm 1,
\mp 1,0),\,(0,0,0,1),\,(1,\pm 1,\pm 1,1)\}$,
\item{(b)} {\it for} $k\equiv 1$ (mod 3), $(a,b,c,d)\in\{(0,0,0,0),\,(0,0,0,-1
),\,(1,0,0,0),\,(1,0,0,-1)\}$,
\item{(c)} {\it for} $k\equiv 2$ (mod 3), $(a,b,c,d)\in\{(0,0,0,0),(-1,\pm 1,
\pm 1,-1),(-1,0,0,0),(0,\pm 1,\mp 1,-1)\}$.\medskip

\noindent{{\it Proof}}\quad First, it is necessary to verify that all the
quadruples $(a,b,c,d)$ listed in Claim 4 give rise to automorphism invariants.
It suffices to verify $T$-invariance and that $S^{(k,k+1)}_{\la\mu,\ka\nu}=
S^{(k,k+1)}_{\sigma'(\la\mu),\sigma'(\ka\nu)}$. This is straightforward
using eqs.($2.2b),(2.2c$).

Let $M=M^\sigma$. Since $C\rho=\rho$, Claim 3 tells us $\sigma(\rho,(1,2))=
(A^w\rho, C^uA^x(1,2))$ and $\sigma((1,2),\rho)=(C^vA^y(1,2),A^z\rho)$.
Consider the conjugation $M'$ of $M$ with $u=v=0$. Eqs.(4.5$a$) and $(2.2c$)
give us the following conditions of $w,x,y,z$:
$$\eqalignno{S^{(k)}_{\rho\rho}S^{(k+1)}_{(1,2),(1,2)}=
S^{(k)}_{A^w\rho,A^w\rho}S^{(k+1)}_{A^x(1,2),A^x(1,2)}:&\quad w^2k+x+x^2(k+1)
\equiv 0\ ({\rm mod}\ 3),&\cr
S^{(k)}_{(1,2),(1,2)}S^{(k+1)}_{\rho\rho}=
S^{(k)}_{A^y(1,2),A^y(1,2)}S^{(k+1)}_{A^z\rho,A^z\rho}:&\quad y^2k+y+z^2(k+1)
\equiv 0\ ({\rm mod}\ 3),&(4.11)\cr
S^{(k)}_{\rho,(1,2)}S^{(k+1)}_{(1,2),\rho}=
S^{(k)}_{A^w\rho,A^y(1,2)}S^{(k+1)}_{A^x(1,2),A^z\rho}:&\quad wyk-w
+xz(k+1)-z\equiv 0\ ({\rm mod}\ 3).&\cr}$$
Here we are exploiting the facts that $S^{(\ell)}_{\rho\la}\ne 0$ (see
$(4.6a$)) and $S^{(\ell)}_{(1,2),(1,2)}\ne 0$ (see Lemma 3(b)).
It is easy to find all solutions $(w,x,y,z)$ to (4.11), and to verify that
each one corresponds to some $M^{\sigma'}$ listed in the Claim.

In other words, the automorphism invariant $M^{\sigma''}=M'(M^{\sigma'})^{-1}$
has $\sigma''(\rho,(1,2))=(\rho,(1,2))$ and
$\sigma''((1,2),\rho)=((1,2),\rho)$. By Claim 3 we may write
$\sigma''(\la,\mu)=(C^eA^f\la,C^gA^h\mu)$. Then by  $(4.5a$)
$$\eqalignno{S^{(k)}_{\rho\la}\,S^{(k+1)}_{(1,2)\mu}=&S^{(k)}_{\rho,C^eA^f\la}
\,S^{(k+1)}_{(1,2),C^gA^h\mu}=S^{(k)}_{\rho\la}S^{(k+1)}_{(1,2),C^gA^h\mu},
&(4.12a)\cr
S^{(k)}_{(1,2),\la}\,S^{(k+1)}_{\rho\mu}=&S^{(k)}_{(1,2),C^eA^f\la}\,
S^{(k+1)}_{\rho,C^gA^h\mu}=S^{(k)}_{(1,2),C^eA^f\la}S^{(k+1)}_{\rho\mu}.
&(4.12b)\cr}$$
But Lemma 3(b) and the fact that $S^{(\ell)}_{\rho\nu}\ne 0$ gives us
$\mu=C^gA^h\mu$, $\la=C^eA^f\la$ -- \ie $\sigma''(\la\mu)=(\la\mu)$. Thus
$M'=M^{\sigma'}$. \qquad QED\medskip

Thus, for each $k$ there will be 24 or 16 automorphism invariants. Of these,
the only ones satisfying $(2.11c)$ (so that they correspond to coset
invariants) are: \item{(1)}\quad $k\equiv 0$\ ({\rm mod}\ 3): (0,0,0,0),
$(0,0,0,1)^c$, ${}^c(1,1,-1,0)$, ${}^c( 1,1,1,1)^c$;
\item{(2)}\quad $k\equiv 1\ ({\rm mod}\ 3)$: (0,0,0,0), $(0,0,0,-1)^c$,
${}^c(1,0,0,0)$, ${}^c(1,0,0,-1)^c$;
\item{(3)}\quad $k\equiv 2\ ({\rm mod}\ 3)$: (0,0,0,0), $(0,1,-1,-1)^c$,
${}^c(-1,0,0,0)$, ${}^c(-1,-1,-1,-1)^c$.

\bigskip\bigskip\noindent{{\bf 5. The physical invariants of $su(3)_k\oplus
su(3)_{k+1}$ involving chiral extensions}}\bigskip

In this section we complete the proof of Thm.2.
We will begin by finding all possibilities for the maximally
extended chiral algebras, and then find all
automorphisms of those chiral algebras.

\bigskip\noindent{{\it 5.1 The weights which can couple to} $(\rho\rho)$}
\medskip

In this subsection we begin the
search for all possible chiral algebra extensions by finding the possible
weights $(\la,\mu)\in\p(k)$ which can {\it couple} to $(\rho\rho)$, \ie{} those
$(\la,\mu)$ which satisfy the norm condition ($T$-invariance)
$${2\over k'}+{2\over k'+1}\equiv {\la^2\over k'}+{\mu^2\over k'+1}\sp({\rm
mod}\sp 2),\eqno(5.1a)$$
as well as satisfy the parity rules (4.3) (with $\ka=\nu=\rho$). Any
weights $(\la,\mu)$ for
which $M_{\rho\rho,\la\mu}\neq 0$ for some physical invariant $M$,
must necessarily satisfy (5.1a) and (4.3). We will also impose the condition
$(2.11a$), namely that
$$t(\la)+t(\mu)\equiv 0\sp ({\rm mod}\sp 3).\eqno(5.1b)$$

For arbitrary algebras, this step (\ie{} enumerating the possible
rho-couplings) is quite tedious. Fortunately the hard work has already
been done by \SU, and all we have to do is collect the pieces. Let us begin
by reviewing some observations from \SUR.

For a given physical invariant $M$ of $su(3)_k\oplus su(3)_{k+1}$, write
${\cal P}_L=\{(\la,\mu)\in\p(k)\,|\, M_{\la\mu,\ka\nu}$ $\ne 0$ for some $\ka,
\nu\}$, and ${\cal J}_L=\{A_k^a A_{k+1}^b\,|\, M_{A^a\rho A^b\rho,\rho
\rho}\ne 0\}$. ${\cal P}_R$ and ${\cal J}_R$ can be defined similarly.
${\cal J}_L(\la\mu)$ denotes the orbit, and $\|{\cal J}_L(\la\mu)\|$ its
cardinality. Then we have

\medskip\noindent{\bf Lemma 4}. (a) {\it For each $(\la,\mu)\in\p(k)$, define
$s(\la,\mu)
=\sum_{\ka,\nu}M_{\rho\rho,\ka\nu}\,S^{(k)}_{\la\ka}\,S^{(k+1)}_{\mu\nu}$.
Then each $s(\la,\mu)\ge 0$, and $s(\la,\mu)>0$ iff} $(\la,\mu)\in {\cal P}_L$.

\item{(b)} {\it For any $a,b,c,d$, $M_{A^a\rho A^b
\rho,A^c\rho A^d\rho}=1$ holds iff $M_{A^a\la A^b\mu,A^c\nu A^d\kappa}=M_{\la
\mu,\nu\kappa}$ for all $\la,\mu,\nu,\kappa$, iff $a\,t(\la)+b\,t(\mu)\equiv
c\,t(\nu)+d\,t(\kappa)$ (mod 3) whenever $M_{\la\mu,\nu\kappa}\ne 0$.
In particular,}
$\forall A_k^a A^b_{k+1}\in {\cal J}_L$, $M_{A^a\la A^b\mu,\ka\nu}
=M_{\la\mu,\ka\nu}$.\medskip

The same comments apply to ${\cal P}_R$ and ${\cal J}_R$, of course.
Lemma 4 holds for general $g$, and more generally, any rational conformal
field theory with a centre. It is proven in \GAN\ (see also Lemma 1 in \SUR).
(a) follows from $S^{(k,k+1)}M=MS^{(k,k+1)}$ and the fact that all
$S_{\rho\rho,\ka\nu}$ are  positive.
(b) follows from calculations from $M=S^{(k,k+1)\dag}MS^{(k,k+1)}$, along the
lines of how
$(2.8c$) implied $(2.8a)$ and $(2.8b)$. Note that (b) implies ${\cal J}_{L,R}$
are (Abelian) groups.

A final useful result concerns the eigenvalues of $M$. Write $M$
as the direct sum
$$M=\oplus_i\,M_i=\left(\matrix{M_1&0&\cdots&0\cr 0&M_2&&0\cr \vdots
&&\ddots&\cr 0&0&&M_\alpha\cr}\right)\eqno(5.2)$$
of indecomposable submatrices $M_i$. Each weight $(\la,\mu)\in\p(k)$ will be
`contained'
in one and only one $M_i$. For convenience always choose $M_1$ to contain
$(\rho\rho)$. Because every entry of each $M_i$ is non-negative, then
$M_i$ has a non-negative eigenvalue $r(M_i)$, called the Perron-Frobenius
eigenvalue, with many remarkable properties. The most important is that if
$s$ is any other (possibly complex) eigenvalue of $M_i$ then $|s|\le r(M_i)$.
See \eg \SUR\ for more details. These $M_i$ will usually turn out to be an
$m\times m$ matrix of the form
$$B_{(\ell,m)}=\left(\matrix{\ell&\cdots&\ell\cr\vdots&&\vdots\cr\ell&
\cdots&\ell\cr}\right),\eqno(5.3)$$
for some $\ell,m$. It is easy to see that $r(B_{(\ell,m)})=\ell m$.

Define sets ${\cal R}_L$ and ${\cal R}_R$ by $(\la,\mu)\in {\cal R}_L$
iff $M_{\la\mu,\rho\rho}\ne 0$, and $(\la,\mu)\in {\cal R}_R$ iff
$M_{\rho\rho,\la\mu}\ne 0$. The following result comes from Lemma 3 in \SUR.

\medskip\noindent{\bf Lemma 5.}
(a) {\it Suppose $M$ has $M_1=B_{(1,m)}$ for some $m$. Then for each $i$,
either $M_i=(0)$ or $r(M_i)=m$. Also, for each
$(\la,\mu)\in\p(k)$, $\sum_{\ka\nu} M_{\la\mu,\ka\nu}^2\le m^2/\|\J_L(\la\mu)
\|$.}

\item{(b)} {\it Now suppose ${\cal R}_L=\J_L(\rho\rho)$ and
${\cal R}_R=\J_R(\rho\rho)$. Suppose $M_{\la\mu,\ka\nu}\ne 0$. Then
$M_{\la\mu,\ka\nu}\le \|\J_L\|/\sqrt{\|\J_L(\la\mu)\|\,\|\J_R(\ka\nu)\|}$.
If in addition $(\la,\mu)$ is  not a fixed
point of $\J_L$ (\ie $J\in\J_L$, $J\ne 0$, implies $J(\la,\mu)\ne (\la,\mu)$)
and also ($\ka\nu$) is
not a fixed point of $\J_R$, then $M_{\la\mu,\ka\nu}=1$; moreover,}
$M_{\la\mu,\alpha\beta}\ne 0$ iff $(\alpha\beta)\in\J_R(\ka\nu)$, and
$M_{\alpha\beta,\ka\nu}\ne 0$ iff $(\alpha\beta)\in\J_L(\la,\mu)$. \medskip

We are now ready for the main result of this subsection:

\medskip \noindent{\bf Proposition 1}. \quad {\it Let $M$ be a physical
invariant of $su(3)_k\oplus su(3)_{k+1}$, satisfying (2.11), so that a coset
invariant can be obtained from it. Then $M_{\rho\rho,
\la\mu}=M_{\la\mu,\rho\rho}\in\{0,1\}$, for all $(\la,\mu)\in\p(k)$. Thus
${\cal R}_L={\cal R}_R$ and ${\cal J}_L={\cal J}_R$. Putting $\rho'=({k+1\over
2},{k+1\over 2})$, $\rho''=({k+2\over 2},{k+2\over 2})$, the possibilities for
${\cal R}_L$ are given by:}

\item{(i)} $k\equiv 0$ (mod 3), $k\ne 9,21$: either ${\cal R}_L=\{(\rho
\rho)\}$ or ${\cal R}_L=\{(\rho\rho),\,(A\rho,\rho),\,(A^2\rho,\rho)\}$;

\item{(ii)} $k\equiv 1$ (mod 3), $k\ne 4$: either ${\cal R}_L=\{(\rho,
\rho)\}$ or ${\cal R}_L=\{(\rho,\rho),\,(A\rho,A\rho),$
$(A^2\rho,A^2\rho)\}$;

\item{(iii)} $k\equiv 2$ (mod 3), $k\ne 5,8,20$: either ${\cal R}_L=\{(\rho,
\rho)\}$ or ${\cal R}_L=\{(\rho,\rho),\,(\rho,A\rho),$
$(\rho,A^2\rho)\}$;

\item{(iv)} $k=4$: ${\cal R}_L$ {\it is either given in (ii), or equals}
$\{(\rho\rho),\,(\rho\rho'')\}$ or $\{(A^a\rho,A^a\rho)
,\,(A^a\rho,A^a\rho'')$ $|\,a=0,1,2\}$;

\item{} $k=5$: ${\cal R}_L$ {\it is either given in (iii), or equals}
$\{(\rho\rho),\,(\rho'\rho)\}$ or $\{(\rho,A^a\rho)
,\,(\rho',A^a\rho)\,|\,a=0,1,2\}$;

\item{} $k=8$: ${\cal R}_L$ {\it is either given in (iii), or equals}
$\{(\rho,A^a\rho),\,(\rho,A^a\rho'')\,|\,a=0,1,2\}$;

\item{} $k=9$: ${\cal R}_L$ {\it is either given in (i), or equals}
$\{(A^a\rho,\rho),\,(A^a\rho',\rho)\,|\,a=0,1,2\}$;

\item{} $k=20$: ${\cal R}_L$ {\it is either given in (iii), or equals}
$$\{(\rho,A^a\rho),\,(\rho,A^a\rho''),\,(\rho,A^a(5,5),\,
(\rho,A^a(7,7))\,|\,a=0,1,2\};$$

\item{} $k=21$: ${\cal R}_L$ {\it is either given in (i), or equals}
$$\{(A^a\rho,\rho),\,(A^a\rho',\rho),\,(A^a(5,5),\rho),\,
(A^a(7,7),\rho)\,|\,a=0,1,2\}.$$

\noindent{{\it Proof}}\quad Consider first $k\equiv 0$ (mod 12).
We see from Lemma 2 that $(\la,\mu)\in{\cal R}_L$ means $\la=A^a\rho$ and
$\mu=A^b\rho$ or $\mu=A^b\rho''$, for some $a,b=0,1,2$. Equation $(5.1b$)
forces $b=0$. Put $m_L=\sum_{a=0}^2M_{A^a\rho\rho,\rho\rho}$, $m_L'=
\sum_{a=0}^2M_{A^a\rho\rho'',\rho\rho}$.

Now put $\la=\rho$, $\mu=(1,4)$ in Lemma 4(a):
$$(m_L+m_L')\sin[{2\pi\over k'+1}]+(m_L-m_L')\sin[{8\pi\over
k'+1}]-(m_L+m_L')\sin[{10\pi\over k'+1}]\ge 0.\eqno(5.4)$$
{}From Lemma 4(b) and $M_{\rho\rho,\rho\rho}=1$ we know $m_L=1$ or 3: $m_L=1$
 corresponds to $\J_L=\{A_k^0
\}$; $m_L=3$ corresponds to $\J_L=\{A_k^0,A_k,A_k^2\}$. In either case, (5.4)
forces $m_L> m_L'$ (since $k>9$), while Lemma 4(b)
implies that if $m_L'\ne 0$ then $m_L\le m_L'$. Together these tell us that
$m_L'=0$.

Similar arguments apply to $m_R$ and $m_R'$. That we have ${\cal R}_L={\cal
R}_R$ here, follows from $S^{(k,k+1)}M=MS^{(k,k+1)}$ evaluated at $(\rho\rho,
\rho\rho)$: it says $m_L=m_R$.

The other $k$ are all handled similarly. (See especially Claim 7 in \SUR.)
\qquad QED

\bigskip\noindent{{\it 5.2 Simple current chiral extensions and their
automorphisms}}\medskip

Write ${\cal R}$ for ${\cal R}_L={\cal R}_R$, and $\J$ for $\J_L
=\J_R$. In this subsection we find all $M$ for which $\R\subseteq
({\cal O}^k_0\rho,{\cal O}^{k+1}_0\rho)$. We see from Prop.\ 1 that indeed all
$M$
satisfy that condition, except at the six exceptional levels $k=4,5,8,9,20,21$.

Note that either $\|\J\|=1$ or 3. If it equals 1, then $M$ must be an
automorphism invariant (see Thm.\ 3 in \GA), hence is listed at the end of
Sect.4. The argument for $\|\J\|=3$ will closely
follow that of Sect.4.

Consider first the easiest case: $k\equiv 1$ (mod 3)\ $(k\not=4).$ Then
$\J=\{J^0,J^1, J^2\}$, where $J=A_kA_{k+1}$. Lemma 4(b) tells us that
$\p:=\p_L=\p_R=\{(\la \mu)\in\p(k)\,|\, t(\la)\equiv -t(\mu)\ ({\rm mod}\
3)\}.$ For each $(\la,\mu)\in\p$, define the $J$-orbit $\langle
\la\mu\rangle=\{(\la,\mu),J(\la,\mu),J^2(\la,\mu)\}$, and $$ch_{\lg
\la\mu\rg}=\sum_{(\la'\mu')\in\lg\la\mu\rg} \chi^{(k)}_{\la'}
\chi^{(k+1)}_{\mu'}.\eqno(5.5a)$$

The special thing here is that there are no fixed points of $\J$.
So Lemma 5(b) tells us that there exists a permutation $\sigma$ of the
$J$-orbits $\lg \la\mu\rg\subset \p$ such that the partition function $Z$
associated to $M$ can be written
$$Z=\sum_{\lg\la\mu\rg} ch_{\lg\la\mu\rg}\,ch^*_{\sigma\lg\la\mu\rg}.
\eqno(5.5b)$$
So our task reduces to finding all bijections $\sigma$ such that $(5.5b$)
is a modular invariant. Define a matrix $S^e$ by
$$S^e_{\lg\la\mu\rg,\lg\ka\nu\rg}:=3\,S^{(k)}_{\la\ka}\,S^{(k+1)}_{\mu\nu}.
\eqno(5.6a)$$
Then $S^e$ is unitary and symmetric, and $M$ commutes with $S^{(k,k+1)}$
iff $\sigma$ is a symmetry of $S^e$.
We may formally define ``fusion rules'' by Verlinde's formula:
$$N^e_{\lg\la\mu\rg,\lg\la\mu\rg,\lg\la\mu\rg}:=\sum_{\lg\ka\nu\rg}
{(S^e_{\lg\la\mu\rg,\lg\ka\nu\rg})^3\over S^e_{\lg\rho\rho\rg,
\lg\ka\nu\rg}}=\sum_{\lg\ka\nu\rg}\sum_{a,b=0}^2{(S^{(k)}_{\la,A^a\ka})^3
(S^{(k+1)}_{\mu,A^b\nu})^3\over S^{(k)}_{\rho,A^a\ka}S^{(k+1)}_{\rho,
A^b\nu}}=N^{(k)}_{\la\la\la}N^{(k+1)}_{\mu\mu\mu},\eqno(5.6b)$$
where the second equality arises by using $(5.6a)$ and $(2.2c$) (the cube
cancels the
extra phases which appear). $\sigma$ will also be a symmetry of these $N^e$.

Either $k'$ or $k'+1$ will be odd -- that one will be coprime to 6 for
$k\equiv1\ ({\rm mod}\ 3).$ Say $k'$ is odd. Write
$\sigma\lg\la\mu\rg=\lg\la'\mu'\rg$. Then Lemma 1 says $\la'\in \O\la$. As in
the proof of Claim 3, eqs.(5.6) imply eqs.(4.7), so Lemma 3(a) gives us
$\mu'\in\O\mu$.

Now $C\rho=\rho$, so $\sigma\lg \rho(1,4)\rg$ must equal $\lg \rho,C^aA^b(1,4)
\rg$ for some $a,b$; but $t(C^aA^b(1,4))\equiv -t(\rho)\ ({\rm mod}\ 3)$, so
$b=0$.  Conjugating if necessary (see eq.($4.10a$)), we may suppose
$\sigma\lg\rho (1,4)\rg=\lg \rho(1,4)\rg$. Similarly, conjugating if necessary,
we may suppose $\sigma\lg (1,4)\rho\rg=\lg(1,4)\rho\rg$. Write $\sigma\lg \la
\mu\rg=\lg C^wA^x\la,C^yA^z\mu\rg$; $(5.6a$) gives us $S^{(k+1)}_{(14),\mu}=
S^{(k+1)}_{(14),C^yA^z\mu}$, $S^{(k)}_{(14),\la}=S^{(k)}_{(14),C^wA^x\la}$.
Lemma 3(c) then requires $\mu=A^i\mu'$ and $\la=A^j\la'$ for some $i,j$.
$t(\la)+t(\mu)\equiv t(\la')+t(\mu')\equiv 0\ ({\rm mod}\ 3)$ then forces
$i=j$, so $\sigma\lg\la\mu\rg =\lg\la\mu\rg$.

The final result (reintroducing the conjugations and ensuring $\sigma\lg \la
\mu\rg\in\p$) is that either $\sigma\lg\la\mu\rg=\lg\la\mu\rg$
$\forall\lg\la\mu\rg$, or $\sigma\lg\la\mu\rg=\lg C\la,C\mu\rg$
$\forall\lg\la\mu\rg$.

The arguments for $k\equiv 0$ and $k\equiv 2$ (mod 3) ($k\not= 5,8,9,20,21$)
are identical  to each other, and similar to $k\equiv 1\ ({\rm mod}\ 3)$ except
for the presence of the fixed points $((k'/3,k'/3),\mu)$ and
$(\la,([k'+1]/3,[k'+1]/3))$, respectively. Consider the case
$k\equiv 0\ ({\rm mod}\ 3)$. Here,
$\p:={\cal P}_L={\cal P}_R=\{(\la,\mu)\in \p(k)\,|\,
t(\la)\equiv 0\ ({\rm mod}\ 3)\}$. Write
$f=(k'/3,k'/3)$, $\lg\la\rg=\{\la, A_k\la,A_k^2\la\}$,
$$ch_{\lg\la\rg\mu}=\sum_{\la'\in\lg\la\rg}\chi_{\la'}^{(k)}\,
\chi_{\mu}^{(k+1)}.\eqno(5.7a)$$
Then as before,
$$Z=\sum M^e_{\lg\la\rg\mu,\lg\ka\rg\nu}\,ch_{\lg\la\rg\mu}\,
ch_{\lg\ka\rg\nu}^*.\eqno(5.7b)$$

Choose any $(\la,\mu)\in\p$, $\la\ne f$, such that $M_{\la\mu,f\nu}=0$ for all
$\nu$. Then Lemma 5(b) says there exists a map $\sigma$ such that
$$M^e_{\lg\la\rg\mu,\lg\ka\rg\nu}=\delta_{\sigma(\lg\la\rg\mu),\lg\ka
\rg\nu}.\eqno(5.8a)$$
Suppose we have $\sigma(\lg\la\rg\mu)=\lg\la'\rg\mu'$ and
$\sigma(\lg\ka\rg\nu)=\lg\ka'\rg\nu'$. Look at $S^{(k,k+1)}M=MS^{(k,k+1)}$, we
get an equation resembling $(4.5)$:
$$S^{(k)}_{\la\ka}\,S^{(k+1)}_{\mu\nu}=S^{(k)}_{\la'\ka'}\,S^{(k+1)}_{\mu'
\nu'}.\eqno(5.8b)$$
By Lemma 2, $M_{\rho\mu,f\nu}=0$ for all $\mu,\nu$, so we may write
$\sigma(\lg\rho\rg\mu)=\lg\la'\rg\mu'$. Lemma 2 tells us that $\la'\in
\lg\rho\rg$ or $\lg\rho'=({k+1\over 2},{k+1\over 2})\rg$ (when $k=9,15,21,57$
we have additional possibilities, but they all succumb to similar arguments).

Suppose for contradiction that $\la'=\rho'$. Apply $(5.8b$) with $\nu=\kappa
=\rho$: we find (see $(4.6c)$) that $S^{(k)}_{\rho\rho'}/S^{(k)}_{\rho\rho}$
must be a rational number. But we proved in Claim 3 that this could only
happen for $k'\le 6$, yet Lemma 2 and $k\equiv 0$ (mod 3) tells us $k'\ge 12$.
Therefore $\sigma(\lg\rho\rg\mu)=\lg\rho\rg\mu'$.

$(5.8b$) now tells us (choosing $\la=\ka=\rho$) that $S^{(k+1)}_{\mu\nu}
=S^{(k+1)}_{\mu'\nu'}$. Thus the map $\mu\mapsto \mu'$ defines an
automorphism invariant of $su(3)_{k+1}$, so equals $\Ab_{k+1}$ or
$\Db_{k+1}$ or their conjugations. Multiplying $M$ by $(\Ab_k\ \otimes$
this automorphism invariant) allows us to suppose that $\sigma(\lg\rho\rg\mu)=
\lg\rho\rg\mu$, for all $\mu$.

Now choose any $\la$ with $t(\la)\equiv 0$. Then
$$\sum_{\alpha,\beta}S_{\rho\alpha}^{(k)}\,S^{(k+1)}_{\nu\beta}\,
M_{\alpha\beta,\la\mu}=\sum_{\alpha,\beta}M_{\rho\nu,\alpha\beta}\,
S^{(k)}_{\alpha\la}\,S^{(k+1)}_{\beta\mu}=3\,S^{(k)}_{\rho\la}\,
S^{(k+1)}_{\nu\mu}\eqno(5.9a)$$
holds for any $\nu$. Multiplying this by $S^{(k+1)*}_{\nu\ka}$ for any $\ka$
and summing over $\nu$, we have
$$\sum_{\alpha}S^{(k)}_{\rho\alpha}\,M_{\alpha\kappa,\la\mu}=3\,
S^{(k)}_{\rho\la}\,\delta_{\kappa\mu}.\eqno(5.9b)$$
By positivity, we see from this that $M_{\alpha\beta,\la\mu}\ne 0$ requires
$\beta=\mu$.

The remainder of the argument is as in Thm.3 of \SUR. Namely,
$M_{(2,2)\rho,f\mu}=0$ for all $\mu$ (unless $k=3$, which can be worked out
by $T$-invariance, and $k=9$, which gives us the exceptional $\Eb^{(2)}_9
\ot \Ab_{10}$); from this we get $\sigma(\lg 2,2\rg\rho)
=\lg 2,2\rg\rho$; $M_{f\mu,\la\mu}=0$ for any $\la\ne f$ because otherwise
$S^{(k,k+1)}M=MS^{(k,k+1)}$ evaluated at $(\rho\rho,\la\mu)$ and at
$((2,2)\rho,\la\mu)$ yield
two incompatible equations. Therefore the fixed point behaviour is trivial:
$M_{f\mu,\la\nu}=3\delta_{f\la}\delta_{\mu\nu}$. $\sigma$ can now be extended
to all $\lg\la\rg\mu$ by defining $\sigma(\lg f\rg\mu)=\lg f\rg \mu$. Then
$(5.8b$) holds $\forall \la,\mu,\nu,\kappa\in\p$. Looking at
$N_{\la ff}^{(k)}N^{(k+1)}_{\mu\mu\mu}$, and using Lemma 3(c), completes
the argument.

The final result, reintroducing the automorphisms and imposing $(2.11c)$, is
that $M={\cal M}_k\ot{\cal M}_{k+1}$ where ${\cal M}_k=\Db_k$ or $\Db_k^c$,
and ${\cal M}_{k+1}=\Ab_{k+1}$ or $\Db_{k+1}^c$. Thus
together with Prop.1 and Claim 4, we have proven Thm.2 for all $k$,
except for $k=4,5,8,9,20$ and 21.

\bigskip\noindent{{\it 5.3 The exceptional levels}}
\medskip

In this subsection we complete the proof of Thm.2 by considering the remaining
levels $k=4,5,8,9,20,21$. We know the values of $M_{\la\mu,\rho\rho}$ and
$M_{\rho\rho,\la\mu}$ for these levels (Prop.1); the task is to go from these
to all $M_{\la\mu,\ka\nu}$. We have many tools for doing this: most notably
Lemmas 1,2,4 and 5, as well as direct use of $T$-invariance and the relation
$MS^{(k,k+1)}=S^{(k,k+1)}M$. The arguments for each level are all essentially
the same; we will explicitly give the one for $k=4$. See Thm.3 of \SUR\ for
more details.

Let us begin with $k=4$, and ${\cal R}=\{(\rho,\rho),\,(\rho,\rho'')\}$.
Then by Prop.1, $M_{\rho\rho,\la\mu}=M_{\la\mu,\rho\rho}=0$, for all
$(\la,\mu)\in\p(k)$, except for $M_{\rho\rho,\rho\rho}=M_{\rho\rho,\rho\rho''}=
M_{\rho\rho'',\rho\rho}=1$. Now suppose $M_{\la\mu,\rho\rho''}\ne 0$.
Choose $\ell$ coprime to $3\cdot 7\cdot 8$, such that $\ell\equiv \pm 1$
(mod $3\cdot 7$) and $\ell\equiv 5$ (mod $3\cdot 8$) -- \eg $\ell=125$.
Then $[\ell\rho]_4=\rho$, $[\ell\rho'']_5=\rho$, so by the parity rule $(4.2a$)
we get $M_{\la'\mu',\rho\rho}=M_{\la\mu,\rho\rho''}\ne 0$, where $\la'=[\ell
\la]_4$ and $\mu'=[\ell \mu]_5$. Thus $M_{\la\mu,\rho\rho''}=1$, and
$(\la'\mu')=(\rho \rho)$ or $(\rho\rho'')$, \ie $(\la\mu)=(\rho\rho'')$ or
$(\rho\rho)$.

What we have shown is that $M_1=B_{(1,2)}$ (see (5.3)). By Lemma 4(a),
$\p=P_{++}(su(3),4)
\times (\O_0\rho\cup \O_0(3,3)\cup \O(1,3))$. Equations (4.1) and (4.3) tell
us $M_{\la\mu,\ka\nu}\ne 0$ requires $\ka\in\O\la$ and either $\mu,\nu\in
\O_0\rho\cup\O_0(3,3)$, or $\mu,\nu\in\O(1,3)$. Now, computing $(S^{(k,k+1)}
M)_{\la\mu,\rho\rho}=(MS^{(k,k+1)})_{\la\mu,\rho\rho}$ for any $(\la,\mu)\in\p$
gives us $\sum_{\ka\nu}M_{\la\mu,\ka\nu}=2$, and if $\mu\in\O_0\rho\cup
\O_0(3,3)$ we get further that there exist weights $(\la'\mu'),(\la''\mu'')\in
\p$ such that $\mu'\in\O_0\rho$ and $\mu''\in\O_0(3,3)$, and
$M_{\la\mu,\la'\mu'}=M_{\la\mu,\la''\mu''}=1$.

For each $a$, there exist $b,c$ such that $M_{\rho A^a\rho,A^b\rho A^c\rho}\ne
0$. Then $(2.11a$) and $T$-invariance force $b=0$ and $a=c$. Thus by Lemma
4(b) and $(2.11b$), $M_{\la\mu,\ka\nu}=M_{A^a\la A^b\mu,A^a\ka A^b\nu}$, and
$M_{\la\mu,\ka\nu}\ne 0$ implies $t(\la)\equiv t(\ka)$ and $t(\mu)\equiv
t(\nu)$.

We can now completely determine $M$, once we know the values of $M_{\rho(1,3),
\rho(1,3)}$, $M_{(1,2)\rho,(1,2)\rho}$, $M_{(1,2)\rho,(1,2)(3,3)}$,
$M_{(1,2)(1,3),(1,2)(1,3)}$ and $M_{(1,2)(1,3),(1,2)(4,3)}$. By multiplying
$M$ by $\Db_4^c\ot \Ab_5$ if necessary, we can force $M_{(1,2)\rho,(1,2)\rho}
=1$. The relation $S^{(k,k+1)}M=MS^{(k,k+1)}$ evaluated at certain weights
now forces $M=\Ab_4\ot \Eb_5$.

The other anomolous possibility for ${\cal R}$ when $k=4$ is ${\cal R}=
\{A^a(\rho,\rho),A^a(\rho,\rho'')\,|\,a=0,1,2\}$. Here $\J$ has order 3 and $
M_1=B_{(1,6)}$, but otherwise the argument is very similar to the one just
given.

\bigskip\bigskip\noindent{{\bf 6. Conclusion}}\bigskip

Our main results are already pointed out in the Introduction. We conclude with
a brief discussion of possible future directions of
research.

The correspondence of section 2, between the physical modular invariants of
diagonal coset theories and WZNW tensor products, was the starting point of
this work. We restricted attention here to diagonal cosets without fixed
points, because of the simplicity of their field identifications. The switch of
weights should also yield a correspondence for more general classes of GKO
cosets. We hope to address this in later work. The correspondence might be
helpful in calculating other quantities of
interest  in coset theories, such as correlation functions.

The ultimate goal of any modular invariant classification such as ours is a
deeper understanding of (rational) conformal field theories in general. There
are two ways progress may be made. First, the truths used to complete a
particular classification can be shown valid in more general contexts. Since
coset theories comprise (at least) a large part of all rational conformal
theories, it would be worthwhile to try to prove some of our results for an
arbitrary rational theory, perhaps with a centre. Second, a pattern may
emerge in the completed classifications that can be extended to other
theories. Presumably, such a pattern would be some generalisation of the
famous A-D-E results of \CIZ. Perhaps the connection with \KR\ points the way.

\bigskip\bigskip\noindent{{\bf Acknowledgements}}\bigskip

T.G. would like to
thank Antoine Coste, Quang Ho-Kim, Philippe Ruelle, and Yassen Stanev for
valuable communications, and the IHES for its generosity. M.W. thanks Pierre
Mathieu for helpful discussions.

\bigskip\bigskip\noindent{{\bf References}}\bigskip

\item{1.} Bais, F.\ A., Bouwknegt, P., Surridge, M., Schoutens, K.:
Coset construction for extended Virasoro algebras. Nucl.\ Phys.\ {\bf B304}
371-391 (1988);

\item{} Di Francesco, P., Zuber, J.-B.: SU(N) lattice integrable models and
modular invariants. In: Proceedings of Trieste conference on Recent
Developments in Conformal Field Theories (1989)

\item{2.} B\'egin, L., Mathieu, P., Walton, M.~A.: $\hat{su}(3)_k$ fusion
coefficients. Mod. Phys. Lett. {\bf A7} 3255-3265 (1992)

\item{3.} Bernard, D.: String characters from Kac-Moody automorphisms.
Nucl.\ Phys.\ {\bf B288} 628-648 (1987);

\item{} Altschuler, D., Lacki, J., Zaugg, P.: The affine Weyl group and
modular invariant partition functions. Phys.\ Lett.\ {\bf 205B} 281-284
(1988);

\item{} Felder, G., Gawedzki, K., Kupiainen, A.: Spectra of
Wess-Zumino-Witten models with arbitrary simple groups. Commun.\ Math.\ Phys.\
{\bf 117} 127-158 (1988);

\item{} Ahn, C., Walton, M.\ A.: Spectra of strings on nonsimply-connected
group  manifolds. Phys.\ Lett.\ {\bf 223B} 343-348 (1989)

\item{4.} Cappelli, A., Itzykson, C., Zuber, J.-B.: The A-D-E classification
of $A^{(1)}$ and minimal conformal field theories. Commun.\ Math.\ Phys.\
{\bf 113} 1-26 (1987); Modular invariant
partition functions in two dimensions. Nucl. Phys. {\bf
B280 [FS18]} 445-465 (1987)

\item{5.} Christe, P., Ravanani, F.: $G_N\otimes G_L/G_{N+L}$ conformal
field theories and their modular invariant partition functions.
Int. J. Mod. Phys. A{\bf 4} 897-920 (1989)

\item{6.} Gannon, T.: WZW commutants, lattices, and level-one partition
functions. Nucl. Phys. {\bf B396} 708-736 (1993)

\item{7.} Gannon, T.: Towards a classification of su(2)$\oplus\cdots
\oplus$su(2) modular invariant partition functions. IHES preprint P/94/21
(hep-th/9402074)

\item{8.} Gannon, T.: The classification of affine SU(3) modular
invariant partition functions. Commun.\ Math.\ Phys.\ {\bf 161} 233-264 (1994)

\item{9.} Gannon, T., Ho-Kim, Q.: The low level modular invariant partition
functions of rank-two algebras. Int.\ J.\ Mod.\ Phys.\ (in press)
(hep-th/9304106); The rank four heterotic modular invariant
partition functions. Nucl.\ Phys.\ {\bf B} (in press) (hep-th/9402027)

\item{10.} Gannon, T.: The classification of SU(3) modular invariants
revisited. IHES preprint (hep-th/9404185)

\item{11.} Goddard, P., Kent, A., Olive, D.:  Virasoro algebras and coset
space models. Phys.\ Lett.\ {\bf 152B} 88-92 (1985);
Unitary representations of the Virasoro and Super-Virasoro
algebras. Commun.\ Math.\ Phys.\ {\bf 103} 105-119 (1986)

\item{12.} Intriligator, K.: Bonus symmetry in conformal field theory. Nucl.\
Phys.\ {\bf B332} 541-565 (1990);

\item{} Schellekens, A.\ N., Yankielowicz, S.: Simple currents, modular
invariants and fixed points. Int.\ J.\ Mod.\ Phys.\ {\bf 5A} 2903-2952 (1990)

\item{13.} Kac, V.~G.: Infinite dimensional Lie algebras, 3rd ed.
Cambridge: Cambridge University Press 1990

\item{14.} Kac, V., Peterson, D.:  Infinite dimensional lie algebras, theta
functions, and modular forms. Adv.\ Math.\ {\bf 53} 125-264 (1984)

\item{15.} Kac, V.~G., Wakimoto, M.: Modular and conformal invariance
constraints in representation theory of affine algebras.
Adv.\ Math.\ {\bf 70} 156-236 (1988)

\item{16.} Koblitz, N., Rohrlich, D.:  Simple factors in the Jacobian of
a Fermat curve. Can.\ J.\ Math.\ {\bf XXX} 1183-1205 (1978)

\item{17.} Lemire, F., Patera, J.: Congruence number, a generalization of
SU(3) triality. J.\ Math.\ Phys.\ {\bf 21} 2026-2027 (1980)

\item{18.} Lerche, W., Vafa, C., Warner, N.: Chiral rings in N=2
superconformal theories. Nucl.\ Phys.\ {\bf B324} 427-474 (1989)

\item{19.} Mathieu, P., S\'en\'echal, D., Walton, M. A.: Field identification
in nonunitary diagonal cosets. Int.\ J.\ Mod.\ Phys.\ A{\bf 7} supplement 1B
731-764 (1992)

\item{20.} Moore, G., Seiberg, N.: Taming the conformal zoo. Phys.\ Lett.\
{\bf 220B} 422-430 (1989)

\item{21.} Ravanini, F.: An infinite class of new conformal field
theories with extended algebras. Mod.\ Phys.\ Lett.\ A{\bf 3} 397-412 (1988)

\item{22.} Ruelle, Ph., Thiran, E., Weyers, J.: Implications of
an arithmetical symmetry of the commutant for modular invariants. Nucl.\ Phys.\
{\bf B402} 693-708 (1993)

\item{23.} Schellekens, A.~N., Yankielowicz, S.: Field identification fixed
points in the coset construction. Nucl.\ Phys.\ {\bf B334} 67-102 (1990)

\item{24.} Stanev, Y.: Local extensions of the SU(2)$\times$SU(2) conformal
current algebras. Vienna ESI preprint (April 1994)

\item{25.} Washington, L.~C.: Introduction to Cyclotomic Fields. Springer
1982

\vfill\eject

\centerline{{\bf Table. } {Physical invariants of} $su(3)_k\oplus su(3)_1
/su(3)_{k+1}$}\medskip
$$\vbox{\tabskip=0pt\offinterlineskip
        \def\tablerule{\noalign{\hrule}}
        \halign to 5.75in{
        \strut#&\vrule#\tabskip=0em plus2em &
        \hfil#&\vrule#&\hfil#&\hfil#&
        \hfil#&\vrule#
        \tabskip=0pt\cr\tablerule
        &&\omit\hidewidth{\bf Level}\hidewidth &&
       \omit\hidewidth {\bf Physical invariants}\hidewidth &&&\cr\tablerule
&&$k\equiv 0$ (mod 3)&&\hfill$\Ab_k\ot\Ab_{k+1}$\hfill&\hfill$\Db_k\ot\Ab_{k+1}
$\hfill&\hfill$\Db_k{}^c\ot\Ab_{k+1}$\hfill&\cr
&&&&\hfill$\Ab_k\ot\Db_{k+1}{}^c$\hfill&\hfill$\Db_k\ot\Db_{k+1}{}^c$\hfill
&\hfill$\Db_k{}^c\ot\Db_{k+1}{}^c$\hfill&\cr
&&$k\equiv 1$ (mod 3)&&\hfill$\Ab_k\ot\Ab_{k+1}$\hfill&\hfill${\bar M}(A_k
A_{k+1})$\hfill&\hfill${}^c{\bar M}(A_kA_{k+1})^{c}$\hfill&\cr
&&&&\hfill$\Db_k{}^c\ot\Ab_{k+1}$\hfill&\hfill$\Ab_k\ot\Db_{k+1}{}^c$\hfill
& \hfill$\Db_k{}^c\ot\Db_{k+1}{}^c$\hfill&\cr
&&$k\equiv 2$ (mod 3)&&\hfill$\Ab_k\ot\Ab_{k+1}$\hfill&\hfill$\Ab_k\ot
\Db_{k+1}$\hfill&\hfill$\Ab_k\ot\Db_{k+1}{}^c$\hfill&\cr
&&&&\hfill$\Db_k{}^c\ot\Ab_{k+1}$\hfill&\hfill$\Db_k{}^c\ot\Db_{k+1}$\hfill
&\hfill$\Db_k{}^c\ot\Db_{k+1}{}^c$\hfill&\cr
&&\hfill$k=4$\hfill&&\hfill$\Ab_4\ot\Eb_5$\hfill&\hfill$\Db_4^c\ot\Eb_5$\hfill
&&\cr
&&&&\hfill$(\Ab_4\ot\Eb_5)\,[{\bar M}(A_4A_5)]$\hfill&\hfill$(\Ab_4\ot\Eb_5)\,
{}^c[{\bar M}(A_4A_5)]^{c}$\hfill&&\cr
&&\hfill$k=5$\hfill&&\hfill$\Eb_5\ot\Ab_6$\hfill&\hfill$\Eb_5\ot\Db_6$\hfill&&
\cr
&&\hfill$k=8$\hfill&&\hfill$\Ab_8\ot\Eb_9{}^{(1)}$\hfill&\hfill$\Db_8{}^c\ot
\Eb_9{}^{(1)}$\hfill&\hfill$\Ab_8\ot\Eb_9{}^{(2)}$\hfill&\cr
&&&&\hfill$\Db_8{}^c\ot\Eb_9{}^{(2)}$\hfill&\hfill$\Ab_8\ot\Eb_9{}^{(2)c}$
\hfill&\hfill$\Db_8{}^c\ot\Eb_9{}^{(2)c}$\hfill&\cr
&&\hfill$k=9$\hfill&&\hfill$\Eb_9{}^{(1)}\ot\Ab_{10}$\hfill&\hfill
$\Eb_9{}^{(1)}\ot
\Db_{10}{}^c$\hfill&\hfill$\Eb_9{}^{(2)}\ot\Ab_{10}$\hfill&\cr
&&&&\hfill$\Eb_9{}^{(2)}\ot\Db_{10}{}^c$\hfill&\hfill$\Eb_9{}^{(2)c}\ot
\Ab_{10}$\hfill&\hfill$\Eb_9{}^{(2)c}\ot\Db_{10}{}^c$\hfill&\cr
&&\hfill$k=20$\hfill&&\hfill$\Ab_{20}\ot\Eb_{21}$\hfill&\hfill$\Db_{20}{}^c
\ot\Eb_{21}$\hfill&&\cr
&&\hfill$k=21$\hfill&&\hfill$\Eb_{21}\ot\Ab_{22}$\hfill&\hfill$\Eb_{21}\ot
\Db_{22}{}^c$\hfill&&\cr
\tablerule\noalign{\smallskip}  }}$$
\centerline{Table. This gives the corresponding WZNW invariant -- see}
\centerline{Sect.4.1 for the notation, and Sect.2.3 for the correspondence.}
\centerline{Thm.2 says this list is complete.}

\end